\documentclass[aps,prd,preprint,nofootinbib,superscriptaddress,tightenlines]{revtex4}
\usepackage{amsfonts}
\usepackage{mathrsfs}
\usepackage{amsmath}
\usepackage{amssymb}
\usepackage{multirow}
\usepackage{subcaption} 
\usepackage{epsfig}
\usepackage{graphicx}
\usepackage{booktabs}
\usepackage{array}
\usepackage{tabularx}
\usepackage{slashed}
\usepackage{ulem}
\usepackage{bm}
\usepackage{color}
\usepackage{hyperref}

\newcommand{\bqa}{\begin{eqnarray}}
\newcommand{\eqa}{\end{eqnarray}}
\newcommand{\beq}{\begin{equation}}
\newcommand{\eeq}{\end{equation}}
\graphicspath{{fig/}{dia/}} \DeclareGraphicsExtensions{.eps}
\begin{document}
\baselineskip 20pt
%%%%%%%%%%%%%%%%%%%%%%%%%%%%%%%%%%%%%%%%%%%%%%%%%%%%%%%%%%%%%%%%%%%%%%
\title{Searching for dark photons in $J/\psi$ decays}

\author{Xiao Liang}
\email{lx321@sdut.edu.cn}
\affiliation{School of Physics and Optoelectronic Engineering, Shandong University of Technology, Zibo, Shandong 255000, China}

\author{Chun-Yuan Li}
\email{lichunyuan@sdut.edu.cn}
\affiliation{School of Physics and Optoelectronic Engineering, Shandong University of Technology, Zibo, Shandong 255000, China}

\author{Bin-Peng Shang}
\email{shangbp@sdut.edu.cn}
\affiliation{School of Physics and Optoelectronic Engineering, Shandong University of Technology, Zibo, Shandong 255000, China}

\author{Zong-Guo Si}
\email{zgsi@sdu.edu.cn}
\affiliation{School of Physics, Shandong University, Jinan, Shandong 250100, China}

\author{Hong-Xin Wang}
\email{wanghx@mail.sdu.edu.cn}
\affiliation{School of Physics, Shandong University, Jinan, Shandong 250100, China}

\author{Xing-Hua Yang}
\email{yangxinghua@sdut.edu.cn}
\affiliation{School of Physics and Optoelectronic Engineering, Shandong University of Technology, Zibo, Shandong 255000, China}

\author{Dai-Xing Zhang}
\email{24412011130@stumail.sdut.edu.cn}
\affiliation{School of Physics and Optoelectronic Engineering, Shandong University of Technology, Zibo, Shandong 255000, China}

%\author{Jun Jiang}
%\email{jiangjun87@sdu.edu.cn}
%\affiliation{School of Physics, Shandong University, Jinan, Shandong 250100, China}

\begin{abstract}

A dark photon is an Abelian gauge boson from a new $U(1)_D$ gauge symmetry, coupled to the Standard Model via kinetic mixing, with $\epsilon$ inducing an effective coupling to the electromagnetic current and $g_\chi$ to a stable dark matter particle $\chi$. We study $J/\psi$ two-body and four-body decays via a light-mass dark photon ($m_U < 3.0$ GeV) in the framework of non-relativistic QCD (NRQCD), considering both visible and invisible decays of the dark photon into SM fermions or dark sector particles.
We investigate the detection sensitivity to the dark photon mass $m_U$ and kinetic mixing parameter $\epsilon$ at both the BESIII and STCF experiments. 

Our results show that, for two-body final states with $m_U<2m_\chi$, BESIII gives $\epsilon$ limits of $9.3\times10^{-4}$ and $7.6\times10^{-4}$ for lepton-pair and hadron signals, while STCF yields $3.7\times10^{-4}$ and $3.1\times10^{-4}$. For invisible decays with $m_U\ge 2m_\chi$, BESIII obtains $1.4\times10^{-3}$ in $0.3\sim0.8\ \text{GeV}$ and STCF reaches $2.3\times10^{-4}$ in $0.3\sim1.9\ \text{GeV}$, with no signals in other mass regions, and visible decays are all severely suppressed. For four-body channels, BESIII give sensitivity $\epsilon<7.6\times10^{-5}$ for $m_U<2.2\ \text{GeV}$, whereas STCF achieves $1.2\times10^{-5}$ across the full mass range. When $m_U\ge 2m_\chi$, visible modes are nearly excluded; BESIII and STCF set invisible-decay limits of $8.8\times10^{-5}$ at masses below $2.4 \ {\rm GeV}$ and $1.4\times10^{-5}$ at masses below $2.8 \ {\rm GeV}$, respectively, with no detectable signals at higher masses. Except for the number $9.3\times10^{-4}$, all the above limits on $\epsilon$ are basically not excluded by collider experiments. Compared with the constraints from two-body final state processes, the limits derived from four-body decay channels lie well below the existing experimental bounds, providing supportive references for constraining this parameter in BESIII and STCF experiments.
Numerical results for the decay ratios $\Gamma/\Gamma_{J/\psi}$ and expected event numbers are presented, along with the significance $S/\sqrt{B}$ and $p_T$ distributions where applicable.

\vspace {2mm} 
%\noindent {PACS number(s): 12.38.Bx, 12.39.Jh, 14.40.Pq, 14.70.Bh}
\end{abstract}

\maketitle

%%%%%%%%%%%%%%%%%%%%%%%%%%%%%%%%%%%%%%%%%%%%%%%%%%%%%%%%%%%%%%%%%%%%
\section{INTRODUCTION}
\label{sec:introduction}
%%%%%%%%%%%%%%%%%%%%%%%%%%%%%%%%%%%%%%%%%%%%%%%%%%%%%%%%%%%%%%%%%%%%

Observations of the cosmic microwave background, large-scale structure, and other cosmological probes imply that roughly one quarter of the total energy density of the Universe resides in a non-luminous, non-baryonic component that interacts predominantly via gravity \cite{Planck:2018vyg}. This makes dark matter (DM) an essential ingredient of the standard cosmological model, even though its nature remains unexplained in the Standard Model (SM) of elementary particle physics. Despite these extensive efforts, searches to date have yielded only null or non-reproducible results.
A minimal and well-motivated framework assumes a simple $U(1)_D$ gauge symmetry for the dark sector. The associated gauge boson couples to the SM sector through kinetic mixing with the $U(1)_Y$ hypercharge symmetry, effectively introducing a feeble interaction between the dark sector and the SM \cite{Alexander:2016aln,Battaglieri:2017aum,Agrawal:2021dbo,Yuan:2022nmu}.
For dark photon searches, there are mainly three methods \cite{Alexander:2016aln,Fabbrichesi:2020wbt,Caputo:2026pdw}: a bump hunt in the visible final state invariant mass for dark photon decays into SM particles (BaBar \cite{BaBar:2014zli}, Belle \cite{Jaegle:2015fme}, KLOE \cite{KLOE-2:2015nli}, APEX \cite{APEX:2011dww}, A1 \cite{A1:2011yso}), a missing energy/momentum search for decays into invisible dark matter (NA64 \cite{NA64:2017vtt,NA64:2023wbi}, BaBar \cite{BaBar:2017tiz}, Belle II \cite{Belle-II:2022jyy}, LDMX \cite{LDMX:2018cma}), and displaced vertex detection for long-lived dark photons (HPS \cite{HPS:2018xkw}, LHCb \cite{Ilten:2015hya,Ilten:2016tkc}). In this work, we focus on the first two approaches.
The theoretical allowed range of the kinetic mixing parameter $\epsilon$ is $10^{-7}$–$10^{-2}$, which is consistent with precision electroweak data and remains accessible to dedicated experimental searches. Meanwhile, ongoing improvements in experimental precision have gradually constrained the parameter space in the $(m_U,\epsilon^2)$ plane, among which the results from BaBar \cite{BaBar:2009lbr,BaBar:2014zli,BaBar:2012bkw}, LHCb \cite{LHCb:2017trq,LHCb:2019vmc}, Belle \cite{Jaegle:2015fme}, BelleII \cite{Belle-II:2022jyy}, KLOE \cite{KLOE-2:2015nli,KLOE-2:2014qxg,KLOE-2:2012lii,KLOE-2:2018kqf} and CMS \cite{CMS:2023hwl} represent the most rigorous exclusions. In the dark photon mass range considered in this work, BaBar finds no significant excess from $0.02 \ {\rm GeV} $ to $ 10.2 \ {\rm GeV}$, and sets upper limits on the mixing strength at the $10^{-4}$-$10^{-3}$ level \cite{BaBar:2014zli}. The Belle Collaboration searches for dark photon and dark Higgs boson in the mass ranges $0.1$–$3.5\ \text{GeV}/c^2$ and $0.2$–$10.5\ \text{GeV}/c^2$, and excludes mixing parameter values above $8\times 10^{-4}$ \cite{Jaegle:2015fme}.

To date, several experiments have searched for dark matter via hadronic decays. For example, LHC propose a search for dark photons using the charm meson decay \cite{Ilten:2015hya}. In this work, we study the dark sector in decay of $J/\psi$ mediated by a dark photon, where $J/\psi$ is a charmonium state composed of a $c\bar{c}$ quark-antiquark pair. Heavy quarkonium systems provide an exceptionally clean environment for probing dark-sector interactions with heavy quarks. The decay of a heavy quark-antiquark bound state is sensitive to dark-sector particles of arbitrarily low mass, making radiative quarkonium transitions a key probe of sub-GeV dark matter \cite{ParticleDataGroup:2024cfk}. Moreover, extensive experimental data on $J/\psi$ decays are currently available, including $8.774\times 10^{10}$ $J/\psi$ events accumulated by the BESIII experiment \cite{BESIII:2021ocn}, and the future Super Tau-Charm Facility (STCF) is designed to achieve $3.4\times10^{12}$ $J/\psi$ events per year \cite{Lyu:2021tlb}. Consequently, $J/\psi$ radiative decays provide a realistic and powerful platform for exploring invisible dark-sector signatures. 
In this paper, we study $J/\psi$ decays into two-body and four-body final states mediated by a dark photon, aiming to investigate the dark sector.
For the hadronization of $J/\psi$, we adopt the nonrelativistic QCD (NRQCD) factorization formalism developed by Bodwin, Braaten, and Lepage \cite{Bodwin:1994jh,Petrelli:1997ge}, which is now widely used in phenomenological studies of quarkonium production and decay. Under the NRQCD factorization formalism, quarkonium production can be separated into short-distance coefficients and long-distance matrix elements. The short-distance coefficients describe the perturbative contribution of the heavy quark pair, while the long-distance matrix elements characterize the nonperturbative hadronization of the heavy quark pair into quarkonium.

Up to now, a certain amount of progress has been made in dark sector phenomenological research, employing various processes as documented in the existing literature.
For instance, dark photon searches using $e^+e^- \to \gamma A^\prime \to \gamma \mu^+\mu^-$ at future $e^+e^-$ colliders (CEPC, ILC, and FCC-ee) are studied in Ref.\cite{He:2017zzr}.
The double dark photon production $e^+e^- \to A^\prime A^\prime $ with $A^\prime$ decaying into a muon pair is also studied at future $e^+e^-$ colliders (CEPC, FCC-ee, ILC/ILD, and IDEA) \cite{Park:2023ygi}.
Based on a parametrised simulation of the IDEA detector, Ref.\cite{Polesello:2025cbi} studied the production of a dark photon at the proposed CERN FCC-ee collider.
In Ref.\cite{Jiang:2019hfn}, our group discussed the vector dark photon and the scalar mediator produced in the processes $e^+e^- \to q\bar{q}A^\prime$ via electron-positron collision.
Ref.\cite{Cheung:2025kmc} investigate the sensitivity of CEPC and FCC-ee at 240 GeV to long-lived dark photons above 2 GeV pair-produced via a light scalar mixing with the Higgs.
The search for massless dark photons via the decays of charmed pseudoscalar-mesons $D^+$, $D^0$ and $D_s^+$, and singly charmed baryons $\Lambda_c^+$, $\Xi_c^+$ and $\Xi_c^0$ is  investigated in Ref.\cite{Su:2020yze}.
Ref.\cite{Su:2019ipw} explored massless dark photons via two-body hyperon decays channel.
Ref.\cite{Gabrielli:2016cut} analyzed a FCNC processes $f \to f^\prime \gamma^\prime$, which is a fermion $f$ decays into a lighter fermion $f^\prime$ plus a massless dark photon.
In addition, the phenomenology of the dark Z boson at International Linear Collider are studied \cite{San:2022uud}.
In this manuscript, we calculate the two-body and four-body decay processes of $J/\psi$ mediated by a light-mass dark photon $(m_U < 3.0 \; {\rm GeV})$. We discuss both the decays of the dark photon into visible Standard Model particles and its decays into invisible dark sector particles, including Dirac fermion, Majorana fermion, and complex scalar particles. 
To make our work more helpful for corresponding experimental studies, we focus on the numerical results for the production decay ratios $\Gamma/\Gamma_{J/\psi}$ and the expected event numbers at BESIII. In addition, the significance $S/\sqrt{B}$ and the transverse momentum $p_T$ distributions are studied where appropriate.

The rest of the paper is organized as follows. 
In Section  \ref{sec:formulation},  
the main theoretical framework of the dark sector and $J/\psi$ production are presented.
In Section \ref{sec:data}, our comprehensive numerical results are discussed.
Section \ref{sec:summary} is reserved for a summary.

%%%%%%%%%%%%%%%%%%%%%%%%%%%%%%%%%%%%%%%%%%%%%%%%%%%%%%%%%%%%%%%%%%%%
\section{FORMULATION}
\label{sec:formulation}
%%%%%%%%%%%%%%%%%%%%%%%%%%%%%%%%%%%%%%%%%%%%%%%%%%%%%%%%%%%%%%%%%%%%
\subsection{$J/\psi$ decay}

Within the NRQCD framework, the differential decay rate of the process $J/\psi\to l^+l^-U $ can be factorized into the short-distance coefficients and the long-distance matrix elements. 
Short-distance coefficients can be obtained perturbatively through Feynman diagram calculations. The Feynman diagrams and their amplitudes can be generated by FeynArts \cite{Hahn:2000kx}. Subsequently, we employ FeynCalc \cite{Shtabovenko:2020gxv} for the evaluation of Dirac traces and Lorentz tensor algebra. 
At the leading order of relative velocity expansion, the projection of a heavy quark pair $(c\bar{c})$ onto $J/\psi$ can be performed by the simple replacement \cite{Bodwin:1994jh,Petrelli:1997ge},
\begin{equation}
       \Pi_p \ = \  \epsilon_\alpha \frac{1}{2\sqrt{M}} \gamma^{\alpha} (\slashed{p} + M)\otimes \frac{\delta_{ij}}{\sqrt{N_{c}}} \ ,
       \label{eq:projection}
\end{equation}
where $p$ and $\epsilon_\alpha$ are the momentum and polarization vector of the $J/\psi$, respectively. $\delta_{ij}$ ensures the color-singlet nature of the $J/\psi$, and $N_c=3$ is the number of quark colors. When evaluating the squared amplitudes $\big|{\cal M}\big|^{2}$,
we need to sum over the polarization vectors of the $J/\psi$, which is given by \cite{Petrelli:1997ge}
\begin{equation}
    \begin{split} 
\sum \epsilon_\alpha \epsilon_{\alpha'}=\Pi_{\alpha \alpha'}=-g^{\alpha \alpha'}+\frac{p^\alpha p^{\alpha'}}{M^2} \ .
    \end{split}
    \label{summation of polariazations}
\end{equation}

While long-distance matrix elements are non-perturbative and expressed in terms of universal parameters. The long-distance matrix elements are determined by the squared Schr$\mathrm{\ddot{o}}$dinger wavefunction at the origin $|\Psi(r=0)|^2$, which is related to the radial wavefunction $R_n(r=0)$,
\begin{equation}
    \begin{split} 
\langle 0|\mathcal{O}^{J/\psi}|0\rangle = |\Psi_{J/\psi}(r=0)|^2 = \frac{|R_{J/\psi}(r=0)|^2}{4 \pi } \ ,
    \end{split}
\end{equation}
Such matrix elements can be determined using approaches including lattice QCD calculations \cite{Bodwin:1996tg}, potential models \cite{Eichten:1995ch}, and experimental measurements. In the present work, we adopt the experimental extraction approach to get the value of $|R_{J/\psi}(r=0)|^2$.

\subsection{ Lagrangian of dark photon }
A compelling framework for connecting a dark sector to the Standard Model (SM) is provided by renormalizable portals \cite{Alexander:2016aln,Battaglieri:2017aum,Agrawal:2021dbo,Yuan:2022nmu}. These correspond to dimension-four operators that couple SM fields to new gauge-singlet degrees of freedom. In this work, we focus on the vector portal, in which a new Abelian $U(1)_D$ gauge symmetry kinetically mixes with the hypercharge $U(1)_Y$ \cite{Alexander:2016aln,Holdom:1985ag,Batell:2009di,Essig:2013lka}. The Lagrangian of relevant gauge terms is
\begin{equation}
    \begin{split} 
\mathcal{L} \supset -\frac{1}{4} \hat{B}_{\mu\nu} \hat{B}^{\mu\nu} - \frac{1}{4} \hat{Z}_{D\mu\nu} \hat{Z}_{D}^{\mu\nu} + \frac{1}{2} \frac{\epsilon}{\cos\theta} \hat{Z}_{D\mu\nu} \hat{B}^{\mu\nu} + \frac{1}{2} m_{D,0}^2 \hat{Z}_D^\mu \hat{Z}_{D\mu}.
    \end{split}
\end{equation}
Here $\hat{B}_{\mu\nu}$ and $\hat{Z}_{D\mu\nu}$ denote the field strength tensors of $U(1)_Y$ and $U(1)_D$, respectively. $\theta$ is the Weinberg mixing angle, and $\epsilon$ is the kinetic mixing parameter.
After electroweak symmetry breaking, the dark photon mixes with two neutral gauge bosons——the photon and the $Z^0$. 
The interaction Lagrangian term of photon and dark photon is $\mathcal{L} \supset \frac{\epsilon}{2} B_{\mu\nu} Z_D^{\mu\nu}$.
In the following, we use $A'_\mu$ to denote this dark photon field and $m_U$ to represent its mass.
The leading-order couplings of the dark photon to fermions are proportional to $\epsilon$ and depend on the ratio $m_{U}/m_{Z}$. For $\epsilon \ll 1$ and $m_{U} \ll m_{Z}$, the couplings are photon-like. In our study we restrict the dark photon mass to $m_{U} \le 3\ \text{GeV}$, so that the mixing with the $Z^0$ boson is negligible, and the photon-like coupling approximation is well justified.
After electroweak symmetry breaking and diagonalization of the gauge-kinetic mixing terms, the dominant low-energy effective interaction is given by $\epsilon e\, A'_\mu J^\mu_{\rm EM}$.
Therefore, the Lagrangian for the dark photon coupling to SM particles is 
\begin{equation}
    \begin{split} 
\mathcal{L} \supset \epsilon e\, x_f A'_\mu J^\mu_{\rm EM}  \ ,
    \end{split}
\end{equation}
among which $x_l=-1$, $x_\nu=0$, and $x_q=2/3$ or $x_q=-1/3$ for quarks. The specific forms of the Lagrangian are
\begin{equation}
    \begin{split} 
\mathcal{L} \supset \epsilon e\, x_f A'_\mu J^\mu_{\rm EM} =
 \begin{cases}
 - \epsilon e\, A'_\mu \bar{l} \gamma^\mu l \ , \\[4pt]
 \dfrac{2}{3} \epsilon e\, A'_\mu \bar{q}_i \gamma^\mu q_i & \; \text{$q_i=u,c$} \ , \\[4pt]
 -\dfrac{1}{3} \epsilon e\, A'_\mu \bar{q}_i \gamma^\mu q_i & \; \text{$q_i=d,s,b$} \ .
\end{cases}
    \end{split}
\end{equation}

%%%%%%%%%%%%%%%%%%%%%%%%%%%%%%%%%%%%%%%%%%%%%%%%%%%%%%%%%%%%%%%%%%%%%%%%%%%%%%%%%%%%%%%%%%%%

Under the new Abelian $U(1)_D$ gauge symmetry, the dark photon interacts with the stable dark matter (DM) particle $\chi$ via the dark gauge coupling $g_\chi$. 
\begin{equation}
    \begin{split} 
\mathcal{L} \supset -g_\chi A'_\mu J^\mu_\chi \ .
    \end{split}
\end{equation}
The form of the dark current $J^\mu_\chi$ depends on the spin of the dark-sector state, and can be divided into Dirac fermion dark matter, Majorana fermion dark matter, and Complex scalar dark matter cases. The corresponding Lagrangian has the form \cite{Krnjaic:2025noj}
\begin{equation}
    \begin{split} 
\mathcal{L} \supset -g_\chi A'_\mu J^\mu_\chi =
\begin{cases}
-g_\chi A'_\mu \bar{\chi}_D\gamma^\mu \chi_D & \; \text{Dirac fermion} \ , \\[4pt]
-\dfrac{1}{2}g_\chi A'_\mu\bar{\chi}_M\gamma^\mu \gamma^5 \chi_M & \; \text{Majorana fermion} \ , \\[4pt]
-ig_\chi A'_\mu\left(\varphi^\dagger \partial^\mu \varphi - (\partial^\mu \varphi^\dagger)\varphi\right) & \; \text{complex scalar} \ .
\end{cases}
    \end{split}
\end{equation}
The mass terms for the three types of dark matter in the Lagrangian are
\begin{equation}
    \begin{split} 
\mathcal{L}_{\text{mass}}^X =
\begin{cases}
- m_\chi \bar{\chi}_D\chi_D & \; \text{Dirac fermion} \ , \\[4pt]
- \dfrac{1}{2} m_\chi \bar{\chi}_M\chi_M & \; \text{Majorana fermion} \ , \\[4pt]
- m_\chi^2 \varphi^\dagger \varphi & \; \text{complex scalar} \ .
\end{cases}
    \end{split}
\end{equation}

In this work, we consider all of these dark matter candidates.

%%%%%%%%%%%%%%%%%%%%%%%%%%%%%%%%%%%%%%%%%%%%%%%%%%%%%%%%%%%%%%%%%%%%%%%%%%%%%%%%%%%%%%%%%%%%%%%%%%%%%%%%%%%%%%%%%%%%%%%%%%

\subsection{Two body final state decay of $J/\psi$}
\begin{figure}[!thbp]
    \centering   \includegraphics[width=0.7\textwidth]{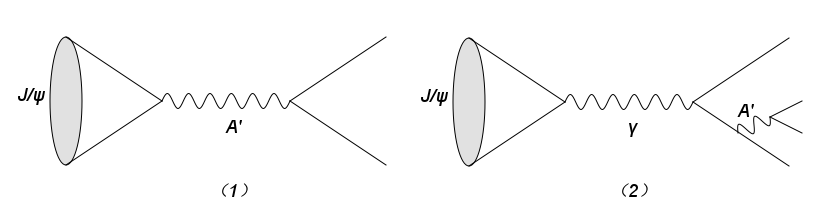}
    \caption{Typical Feynman diagrams for two-body final state decay of $J/\psi$ (left) and four-body final state decay of $J/\psi$ (right) mediated by a dark photon $A'$.} 
    \label{feynman}
\end{figure}

In this work, we discuss the two body final state decay of $J/\psi$ mediated by a dark photon. The typical Feynman diagram is displayed in Fig.~\ref{feynman} panel (1). The decay processes are governed by the visible width corresponding to decays into SM states, as well as the invisible width corresponding to decays into dark states. 
\begin{equation}
    \begin{split} 
\Gamma_{\text{total}} =
\begin{cases}
\Gamma_{\text{vis}} & \; \text{$m_U < 2 m_\chi$} \ , \\[4pt]
\Gamma_{\text{vis}}+\Gamma_{\text{inv}} & \; \text{$m_U \geq 2 m_\chi$} \ .
\end{cases}
    \end{split}
\end{equation}
here $m_U$ is the mass of the dark photon, which can be obtained via either a Stueckelberg term or a dark Higgs mechanism \cite{Fayet:1980ad,Fayet:2004bw,Boehm:2003hm,Pospelov:2007mp,Batell:2009di,Batell:2009yf}. The form of $\Gamma_{\text{vis}}$ becomes\footnote{The explicit expressions for the decay widths $\Gamma(J/\psi \to U \to l^+l^-)$, $\Gamma(J/\psi \to U \to \bar{\chi}_D\chi_D)$, $\Gamma(J/\psi \to U \to \bar{\chi}_M\chi_M)$ and $\Gamma(J/\psi \to U \to \varphi^\dagger \varphi)$ are presented in Appendix~\ref{Explicit Amplitude Form}.}.
\begin{equation}
    \begin{split} 
\Gamma_{\text{vis}}(m_U) = 
\sum_{l=e,\mu} \Gamma(J/\psi \to U \to l^+l^-) + \Gamma(J/\psi \to U \to \text{hadrons}) \ ,
    \end{split}
\end{equation}

The width of dark photon decay to hadrons is incorporated via a global R-ratio,
\begin{equation}
    \begin{split} 
\Gamma (J/\psi \to U \to \text{hadrons}) \ = \ R(\sqrt{s}) \ \Gamma (J/\psi \to U \to \mu^+\mu^-) \ ,
    \end{split}
\end{equation}
where we use the ratio of $R(\sqrt{s})=\sigma_{e^+e^- \to \text{hadrons}} / \sigma_{e^+e^- \to \mu^+\mu^-}$, with $\sqrt{s}$ denoting the center-of-mass invariant energy of the reaction system~\cite{HADES:2013nab,Schmidt:2021hhs}. For the two-body final-state decay, the dark photon is off shell, and thus $R(\sqrt{s})=R(M_{J/\psi})$. Ref.~\cite{KEDR:2018hhr} presents the evaluation of the R ratio in the energy range from 1.84 GeV to 3.72 GeV at the KEDR detector, and we deduce $R(M_{J/\psi})=2.16$ from the corresponding measured results.
The form of $\Gamma_{\text{inv}}$ is
\begin{equation}
    \begin{split} 
\Gamma_{\text{inv}}(m_U) = \sum_{l=\chi_D, \; \chi_M, \; \varphi}
\Gamma(J/\psi \to U \to \bar{\chi}\chi) \ .
    \end{split}
\end{equation}

It should be noted that in the calculation, the total decay width of the dark photon $\Gamma_U$ is approximated as the sum of two-body final state processes. To satisfy this condition and improve the accuracy of our calculation, we need to choose dark photon masses that are well separated from the $2m_e$ and $2m_\mu$ thresholds. Therefore, we select dark photon masses in the range from 0.3 GeV to 3 GeV. 
%%%%%%%%%%%%%%%%%%%%%%%%%%%%%%%%%%%%%%%%%%%%%%%%%%%%%%%%%%%%%%%%%%%%%%%%%%%%%%%%%%%%%%%%%%%%%%%%%%%%%%%%%%%%%%%%%%%%%%%%
\subsection{Four body final state decay of $J/\psi$}

\begin{figure}[!thbp]
    \centering   \includegraphics[width=0.6\textwidth]{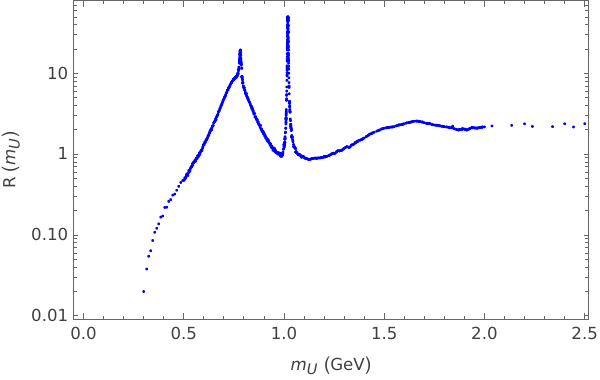}
    \caption{The curves show the parameter $R(m_U)$ as a function of $m_U$ in the range of $ 2m_\mu \leq m_U< M_{J/\psi}$.} 
    \label{R picture}
\end{figure}

We research the four body final state decay $J/\psi \to l^+l^-U \to l^+l^-X$. The typical Feynman diagram is displayed in Fig.~\ref{feynman} panel (2). The decay width of the total decay process can be decomposed into 
\begin{equation}
    \begin{split} 
    \label{15}
\Gamma (J/\psi \to l^+l^-U \to  l^+l^-X ) \ =  \Gamma (J/\psi \to l^+l^-U) \times {\rm Br}(U \to X) \ .
    \end{split}
\end{equation}
Here, the branching ratio ${\rm Br}(U \to X)$ accounts for visible decays ${\rm Br}(U \to e^+e^-)$, ${\rm Br}(U \to \mu^+ \mu^-)$ and ${\rm Br}(U \to \text{hadrons})$, as well as invisible decays ${\rm Br}(U \to \bar{\chi}_D\chi_D)$, ${\rm Br}(U \to \bar{\chi}_M\chi_M)$ and ${\rm Br}(U \to \varphi^\dagger \varphi)$. It takes the form
\begin{equation}
    \begin{split} 
{\rm Br}(U \to X) = 
\begin{cases}
\renewcommand{\arraystretch}{1.5}
\dfrac{\Gamma(U \to X)}{\Gamma _{\text{vis}}}  & \; \text{$m_U < 2 m_\chi$} \ , \\[12pt]
\dfrac{\Gamma(U \to X)}{\Gamma _{\text{vis}} +\Gamma_{\text{inv}}}  & \; \text{$m_U \geq 2 m_\chi$} \ .
    \end{cases}
    \end{split}
\end{equation}
Here $\Gamma_{\text{vis}}$ and $\Gamma_{\text{inv}}$ become \footnote{The explicit expressions for the squared amplitude of the process $J/\psi \to l^+l^-U$ and the decay widths $\Gamma(U \to l^+l^-)$, $\Gamma(U \to \bar{\chi}_D\chi_D)$, $\Gamma(U \to \bar{\chi}_M\chi_M)$ and $\Gamma(U \to \varphi^\dagger \varphi)$ are presented in Appendix~\ref{Explicit Amplitude Form}.}
\begin{equation}
    \begin{split} 
&\Gamma_{\text{vis}}(m_U) = 
\sum_{l=e,\mu} \Gamma( U \to l^+l^-) + \Gamma( U \to \text{hadrons}) \ ,  \\
&\Gamma_{\text{inv}}(m_U) = \sum_{l=\chi_D, \; \chi_M, \; \varphi}
\Gamma( U \to \bar{\chi}\chi) \ ,
    \end{split}
\end{equation}
where
\begin{equation}
    \begin{split} 
 \Gamma( U \to \text{hadrons}) = R(\sqrt{s})\  \Gamma( U \to \mu^+\mu^-).
    \end{split}
\end{equation}
For the four-body final-state decay, the dark photon we consider is on shell. Therefore, the center-of-mass energy $\sqrt{s}$ is equivalent to the on-shell dark photon mass $m_U$, leading to $R(\sqrt{s})=R(m_U)$. Using the data from PDG2024~\cite{ParticleDataGroup:2024cfk}, we extract the behavior of $R(m_U)$ in the mass range $2m_\mu \leq m_U < M_{J/\psi}$ and present the results in Fig.~\ref{R picture}.

Therefor, decay branching ratio of the target process is
\begin{equation}
    \begin{split} 
\mathcal{B} (J/\psi \to  l^+l^-X ) \ = \frac{\Gamma (J/\psi \to l^+l^-X)}{\Gamma_{J/\psi}} =  \frac{\Gamma (J/\psi \to l^+l^-U) }{\Gamma_{J/\psi}}\times {\rm Br}(U \to X)  \ ,
    \end{split}
    \label{B}
\end{equation}
 
Since we aim to study the dark photon reconstructed via the invariant mass spectrum of final-state particles, assuming it is produced on-shell without off-shell contributions, the resulting yield is expected to be lower than the directly measured value. Therefore, our prediction can serve as a baseline for future experimental searches.

%%%%%%%%%%%%%%%%%%%%%%%%%%%%%%%%
\section{NUMERICAL RESULTS}
\label{sec:data}
%%%%%%%%%%%%%%%%%%%%%%%%%%%%%%%%

When performing the numerical calculations, the necessary input parameters are taken as follows \cite{ParticleDataGroup:2024cfk}:
\begin{equation*}
    \begin{split} 
 & \alpha=1/137.065,\;  M_{J/\psi}=3.0969 \; {\rm GeV},\; \Gamma_{J/\psi}=92.6 \; {\rm keV},\; \\ &  m_e=0.511 \; {\rm MeV}, \;
m_\mu =0.106 \;{\rm GeV} \;   .
    \end{split}
\end{equation*}
To determine the value of the $J/\psi$ radial wavefunction at the origin, we use the experimental decay branching ratio $\Gamma(J/\psi \to e^+e^-)/\Gamma_{J/\psi}$ and $\Gamma(J/\psi \to \mu^+\mu^-)/\Gamma_{J/\psi}$, and extract the value from their average for our two subsequent numerical evaluations. We obtain
\begin{eqnarray*}
|R_{J/\psi}(0)|^2 = 0.5599 \; {\rm GeV^3} \;   .
\end{eqnarray*}
For dark coupling $\alpha_\chi$, a larger $\alpha_\chi$ leads to a larger dark matter annihilation cross-section, thereby reducing its relic density. In the thermal freeze-out mechanism, $\alpha_\chi$, $\epsilon$, and $m_{U}$ collectively determine the annihilation cross-section to account for the observed dark matter relic density. In this work, we adopt a theoretically reasonable benchmark value $\alpha_\chi = 0.05$, where
\begin{eqnarray*}
\alpha_\chi = \frac{g_\chi^2}{4\pi} = 0.05 \;   .
\end{eqnarray*}

%%%%%%%%%%%%%%%%%%%%%%%%%%%%%%%%%%%%%%%%%%%%%%%%%%%%%%%%%%%%%%%%%%%%%%%%%%%%%%%%%%%%%%%%%%%%%%
\subsection{two-body final-state decay of $J/\psi$ mediated by the dark photon}
\subsubsection*{\rm{(I)} $ m_U < 2m_\chi$}

\begin{figure}[!thbp]
    \centering   \includegraphics[width=0.75\textwidth]{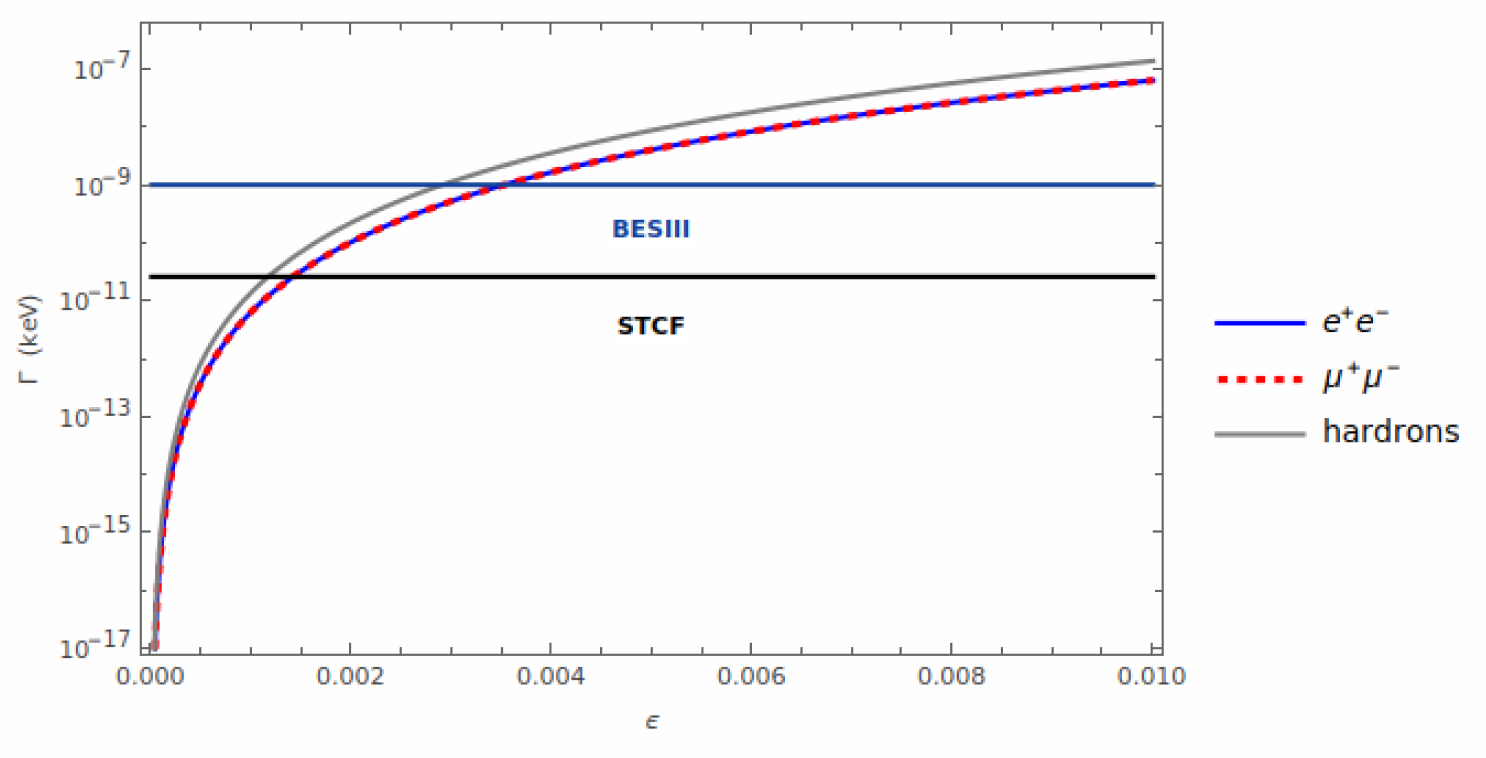}
    \caption{Decay width as a function of the kinetic mixing parameter $\epsilon$ for the visible decay processes $J/\psi \to U \to e^+e^-$, $J/\psi \to U \to \mu^+\mu^-$ and $J/\psi \to U \to \text{hadrons}$, with $m_U=1.0\; {\rm GeV}$. The dark blue and black horizontal lines represent the experimental limits of the BESIII and the STCF, respectively.} 
    \label{epsilon1}
\end{figure}

When $m_U < 2m_\chi$, only visible decays occur. 
In Fig.~\ref{epsilon1}, we display the decay width as a function of the kinetic mixing parameter $\epsilon$ for the visible decay processes $J/\psi \to U \to e^+e^-$, $J/\psi \to U \to \mu^+\mu^-$ and $J/\psi \to U \to \text{hadrons}$ by fixing the dark photon mass $m_U=1.0\; {\rm GeV}$ as an example, as well as the experimental limits for the BESIII and the STCF.
According to the BESIII experiment, the total number of $J/\psi$ events is $8.774 \times 10^{10}$ with a luminosity of about $2568.07\ \text{pb}^{-1}$ \cite{BESIII:2021cxx}, and the STCF experiment is designed to achieve $3.4\times10^{12}$ $J/\psi$ events with an instantaneous luminosity greater than $0.5 \times 10^{35}\ \text{cm}^{-2} \text{s}^{-1}$ \cite{Lyu:2021tlb}. For BESIII, no lepton or hadron signals can be detected when $\epsilon$ is below approximately $9.3 \times 10^{-4}$ and $7.6 \times 10^{-4}$, respectively, while these $\epsilon$ limits can be further lowered to $3.7 \times 10^{-4}$ and $3.1 \times 10^{-4}$ in STCF, for the entire mass range $0.3\ \text{GeV} \leq m_U \leq 3.0\ \text{GeV}$. Our theoretical prediction for the ratio of muon pair production to electron pair production is:
\begin{eqnarray*}
\frac{\Gamma(J/\psi \to U \to \mu^+\mu^-)}{\Gamma(J/\psi \to U \to e^+e^-)} =0.999992  \ .
\end{eqnarray*}

\begin{figure}[!thbp]
    \centering   \includegraphics[width=0.75\textwidth]{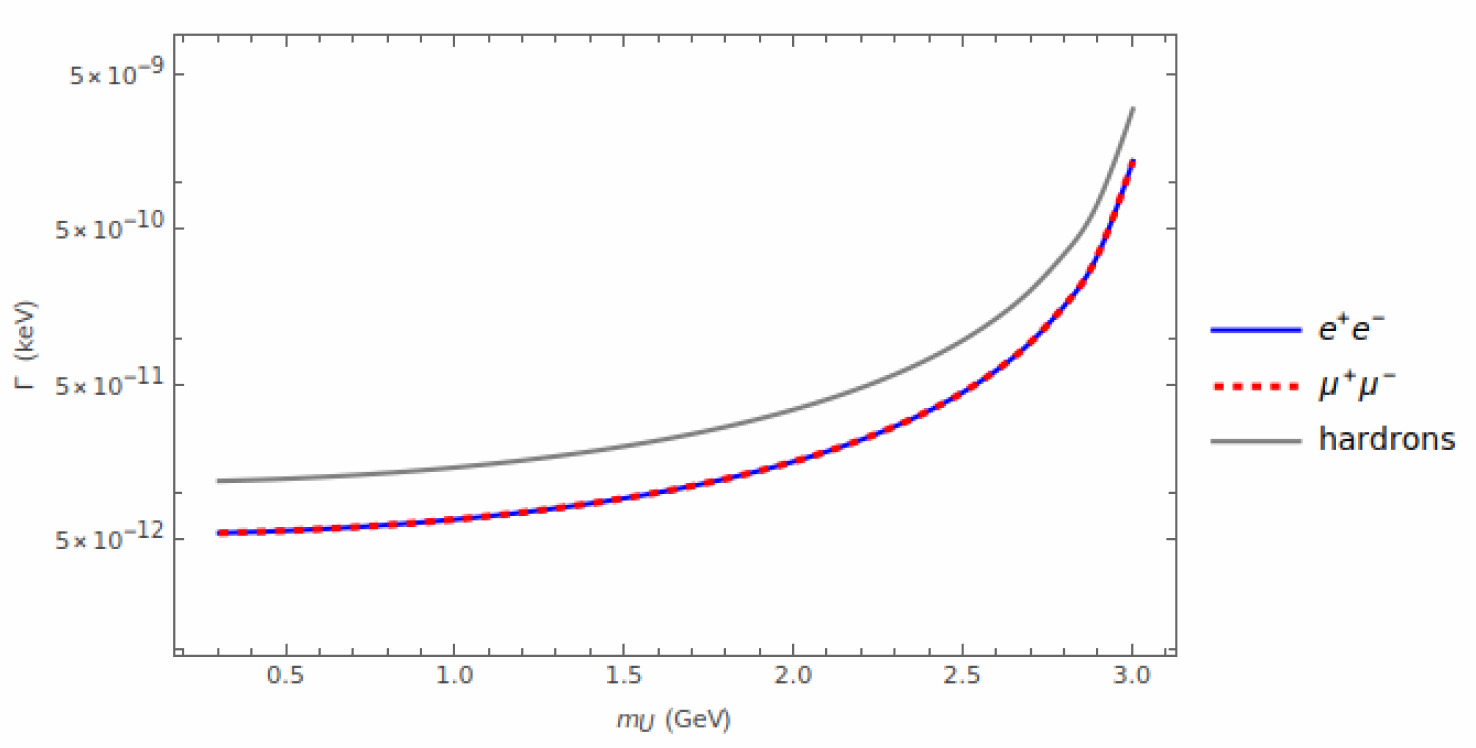}
    \caption{Decay width as a function of the dark photon mass $m_U$ for the visible decay processes $J/\psi \to U \to e^+e^-$, $J/\psi \to U \to \mu^+\mu^-$ and $J/\psi \to U \to \text{hadrons}$, with $\epsilon=10^{-3}$.} 
    \label{twovisiable}
\end{figure}

Then, we fix the coupling parameter $\epsilon = 10^{-3}$ as an example, and perform analyses on the event yields and kinematic distributions at BESIII and STCF. In Fig.~\ref{twovisiable}, we display the decay width as a function of the dark photon mass $m_U$. In addition, we adopt several typical mass values, $m_U=\{0.5, 1.0, 1.5, 2.0, 2.5\ \text{GeV}\}$, and list their corresponding decay ratios $\Gamma/\Gamma_{J/\psi}$ in Table~\ref{two body visible}. The results for all processes show an overall increase with increasing $m_U$. The main reason for such curve characteristics is that there exists a dominant proportional correlation between $\Gamma$ and $1/(m_U^2-M_{J/\psi}^2)^2$. Here we neglect the term $m_U^2\Gamma_U^2$, since $\Gamma_U=\Gamma_{vis}$ is directly proportional to $\epsilon^2$, which makes $m_U^2\Gamma_U^2$ sufficiently small.
As shown in the figure, taking into account the BESIII detection efficiency of $59.21 \%$ and $63.53 \%$ for $J/\psi$ decaying into $e^+e^-$ and $\mu^+\mu^-$, respectively \cite{BESIII:2013csc}, the observed signal yield is almost negligible.  
However, at the STCF, the number of events for the $J/\psi \to e^+e^-$ (and similarly $J/\psi \to \mu^+\mu^-$) and $J/\psi \to \text{hadrons}$ processes mediated by a dark photon is about $0\sim51$ (in mass range $2.3\ \text{GeV} \leq m_U \leq 3.0\ \text{GeV}$) and $0\sim110$ (in mass range $1.8\ \text{GeV} \leq m_U \leq 3.0 \ \text{GeV}$), respectively.
Using the $J/\psi  \to e^+e^-$ decay channel as an example, we estimate the significance $S/\sqrt{B}$ is smaller than $1.132\times10^{-4}$ at the STCF, where the branching ratio of the background process $\mathcal{B}(J/\psi \to e^+e^-)= 5.971 \ \%$ is obtained from PDG2024~\cite{ParticleDataGroup:2024cfk}, which indicates that firm control of statistical and systematic issues is necessary to achieve reliable results. We predict that increasing the event yield of $J/\psi$ by two orders of magnitude will lead to a corresponding increase in all significances by one order of magnitude.
% \cite{Alexander:2016aln}.

 \begin{table}[ht]
    \caption{Predictions for decay ratios $\Gamma/\Gamma_{J/\psi}$ of two body final state visible decay process for $m_U={0.5, \, 1.0, \, 1.5, \, 2.0, \, 2.5, \, 3.0 \; {\rm GeV}}$ with $\epsilon = 10^{-3}$. The theoretical uncertainty entirely propagating from experimental error of $J/\psi$ mass.}
    \begin{center}
    \renewcommand{\arraystretch}{1.4}
      \begin{tabular}{p{2.5cm}<{\centering}| p{3.2cm}<{\centering} |p{3.2cm}<{\centering}| p{3.2cm}<{\centering} | p{3.2cm}<{\centering} }
       \toprule
        \botrule
            $\mathcal{B} \, (\times 10^{-13})$ & $e^+e^-$  & $\mu^+\mu^-$  &  $\text{hadrons}$ & $\Gamma_{vis}/\Gamma_{J/\psi}$   \\
        \hline
$0.5 \; {\rm GeV}$ &  $0.629 \pm 0.0113$ &   $0.629 \pm 0.0113$ & $1.359\pm0.0245$ &  $2.616\pm0.0472$  \\
 \hline
$1.0 \; {\rm GeV}$ &  $0.744 \pm 0.0134$ &  $0.744 \pm 0.0134$ & $1.606\pm0.0290 $ & $3.093\pm0.0558$   \\
 \hline
$1.5 \; {\rm GeV}$ &  $1.0183 \pm 0.0184$ &  $1.0183 \pm 0.0184$ & $2.200\pm0.0397$ & $4.236\pm0.0764$   \\
 \hline
$2.0 \; {\rm GeV}$ &  $1.755 \pm 0.0316$ &  $1.755 \pm 0.0316$ & $3.792\pm0.0684$ & $7.303\pm0.132$   \\    
 \hline
$2.5 \; {\rm GeV}$ &  $4.913 \pm 0.0886$ &  $4.913 \pm 0.0886$ & $10.612\pm0.191$ & $20.437\pm0.368$   \\
        \botrule
      \end{tabular}
    \end{center}
    \label{two body visible}
\end{table}

Finally, we perform a more essential phenomenological analysis for the visible decay process $J/\psi \to U \to e^+e^-$ as an example \footnote{The explicit expression for the differential decay width $d\Gamma(J/\psi \to U \to l^+l^-)/dp_T$ is presented in Appendix~\ref{Explicit Amplitude Form}.}. Fig.~\ref{pt} displays the transverse momentum $p_T$ distributions of the decay width for various dark photon masses. It is found that the distribution rises sharply as $m_U$ increases.
%%%%%%%%%%%%%%%%%%%%%%%%%%%%%%%%%%%%%%%%%%%%%%%%%%%%%%%%%%%%%%%%%%%%%%%%%%%%%%%%%%%%%%%%%%%%%%%%%%%%%%%%%%%%%%%%%%%%%%%%%%%%%%%%%%%%%%%

\subsubsection*{ $ {\rm (II)} \ m_U \geq 2m_\chi$}

\begin{figure}[!thbp]
    \centering   \includegraphics[width=0.49\textwidth]{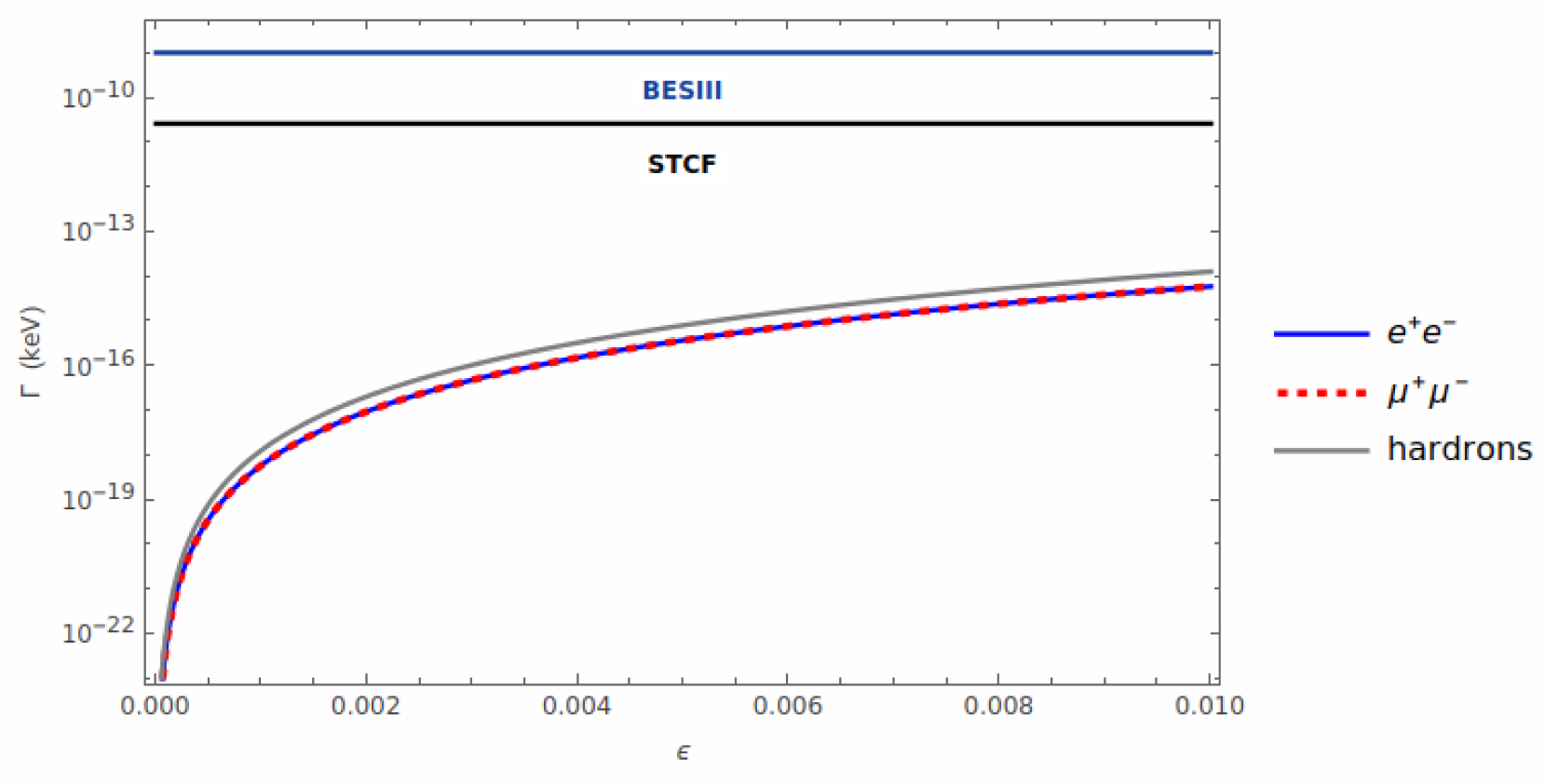} \includegraphics[width=0.47\textwidth]{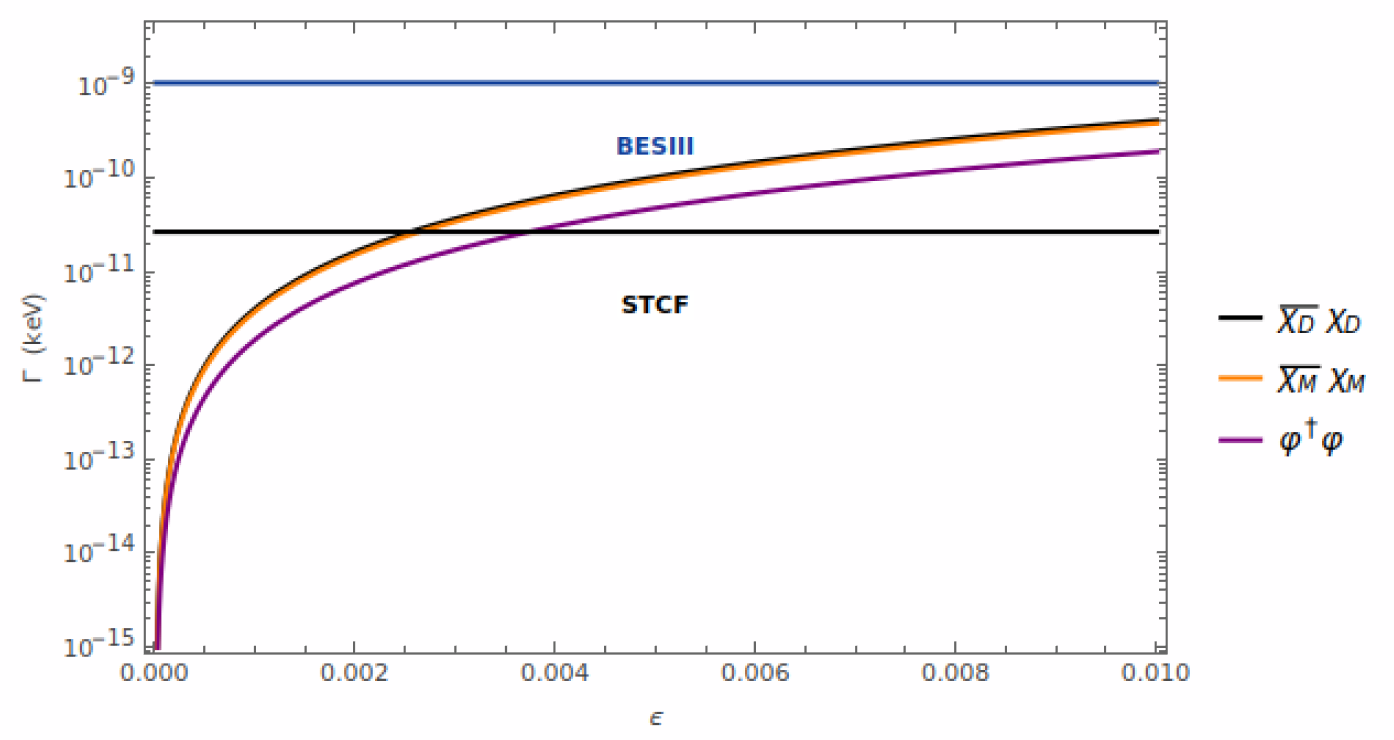}
    \caption{
    Left: Decay width as a function of the kinetic mixing parameter $\epsilon$ for the visible decay processes $J/\psi \to U \to e^+e^-$, $J/\psi \to U \to \mu^+\mu^-$ and $J/\psi \to U \to \text{hadrons}$. The dark blue and black horizontal lines represent the experimental limits of the BESIII and the STCF, respectively. Right: Decay width as a function of the kinetic mixing parameter $\epsilon$ for the invisible decay processes $J/\psi \to U \to \bar{\chi}_D\chi_D$, $J/\psi \to U \to \bar{\chi}_M\chi_M$ and $J/\psi \to U \to \varphi^\dagger \varphi$, with $\alpha_\chi = 0.05$ and $m_U/m_\chi=3.0$. Both figures correspond to $m_U=1.0\; {\rm GeV}$.} 
    \label{epsilon2}
\end{figure}

When $m_U \geq 2m_\chi$, visible and invisible decays occur simultaneously. In the corresponding experiments, the presence of invisible decays may be inferred from a resonance-like feature in the missing-mass or missing-momentum distribution with a single photon. In Fig.~\ref{epsilon2}, we display the decay width as a function of the kinetic mixing parameter $\epsilon$ for the visible decay processes $J/\psi \to U \to e^+e^-$, $J/\psi \to U \to \mu^+\mu^-$ and $J/\psi \to U \to \text{hadrons}$, as well as the invisible decay processes $J/\psi \to U \to \bar{\chi}_D\chi_D$, $J/\psi \to U \to \bar{\chi}_M\chi_M$ and $J/\psi \to U \to \varphi^\dagger \varphi$ by fixing the dark photon mass $m_U=1.0\; {\rm GeV}$ as an example. It can be seen that, within the theoretical range of $\epsilon$ from $10^{-2}$ to $10^{-7}$, all visible decays are excluded in both the BESIII and STCF experiments. This is because the visible final states of SM particles are strongly suppressed by the invisible decay channels when $m_U \geq 2m_\chi$.
For invisible decays, $\epsilon$ limit is $1.4\times10^{-3}$ for the dark photon mass range $0.3\ \text{GeV} \leq m_U \leq 0.8\ \text{GeV}$, but for masses above $0.8\ \text{GeV}$, no signals can be detected within the theoretical allowed range of $\epsilon$ in BESIII. Meanwhile, the $\epsilon$ limit is $2.3\times10^{-4}$ for the dark photon mass range $0.3\ \text{GeV} \leq m_U \leq 1.9\ \text{GeV}$, but for masses above $1.9\ \text{GeV}$, no signals can be detected in STCF.

\begin{figure}[!thbp]
    \centering   \includegraphics[width=0.5\textwidth]{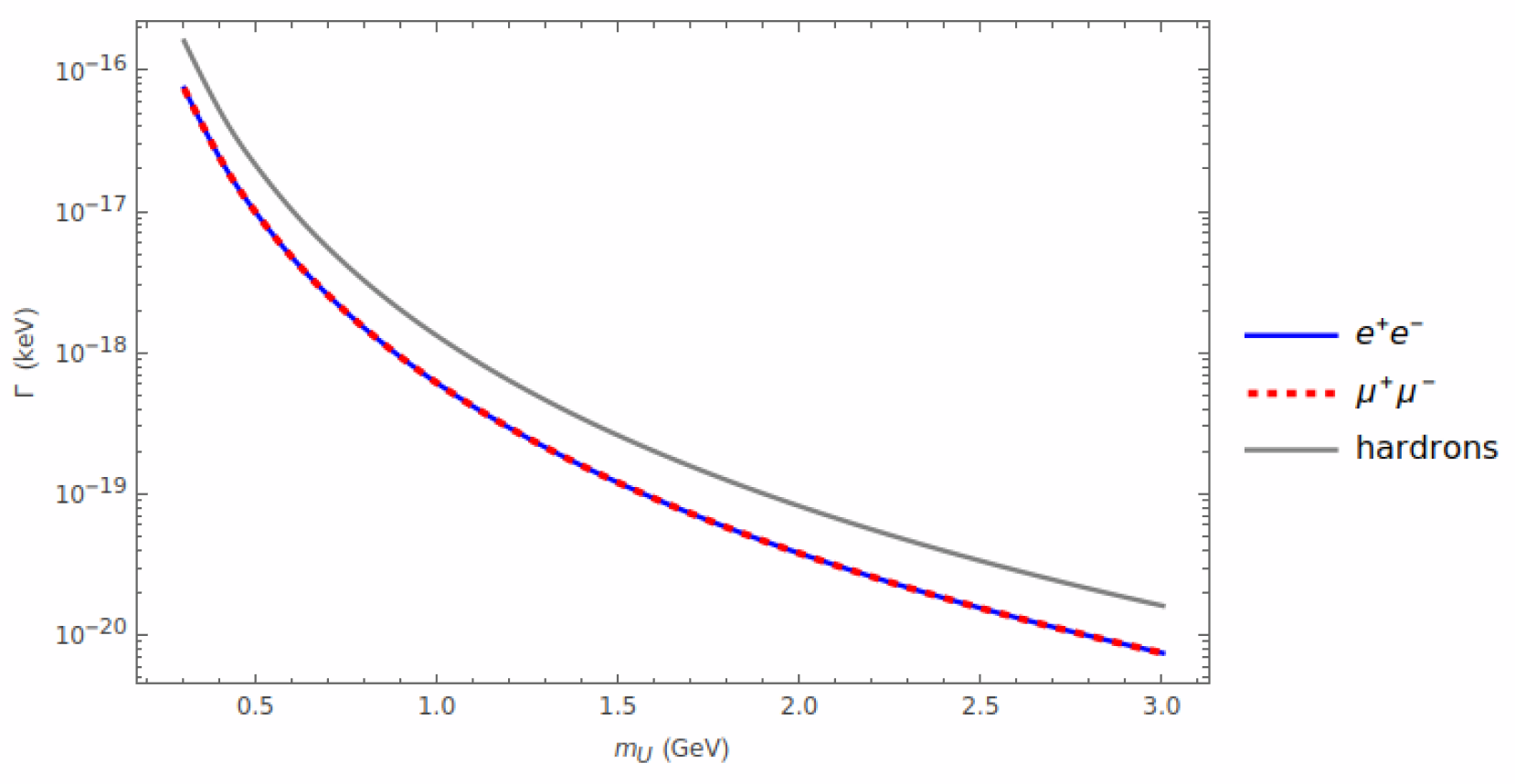} \includegraphics[width=0.47\textwidth]{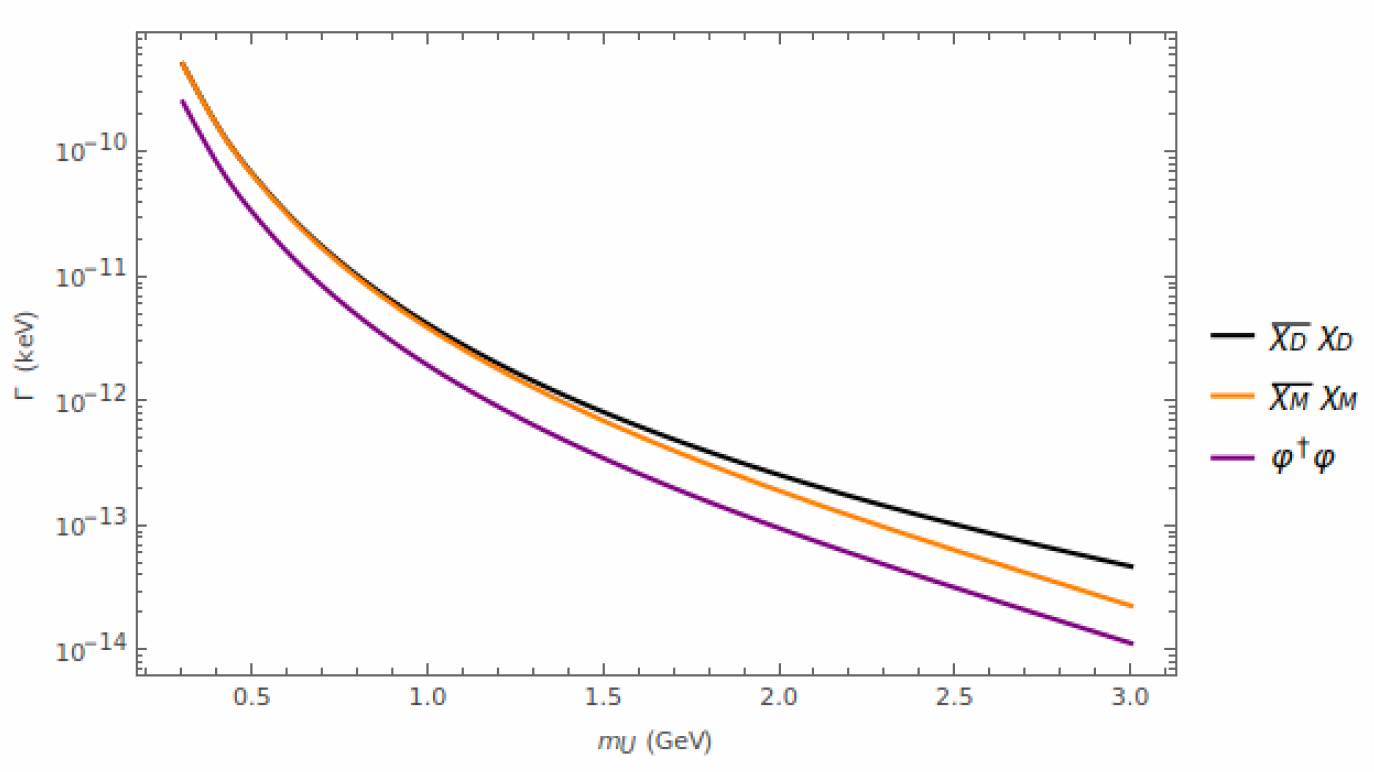}
    \caption{Left: Decay width as a function of the dark photon mass $m_U$ for the visible decay processes $J/\psi \to U \to e^+e^-$, $J/\psi \to U \to \mu^+\mu^-$ and $J/\psi \to U \to \text{hadrons}$. Right: Decay width as a function of the dark photon mass $m_U$ for the invisible decay processes $J/\psi \to U \to \bar{\chi}_D\chi_D$, $J/\psi \to U \to \bar{\chi}_M\chi_M$ and $J/\psi \to U \to \varphi^\dagger \varphi$, with $\alpha_\chi = 0.05$ and $m_U/m_\chi=3.0$. Both figures correspond to $\epsilon=10^{-3}$.}
    \label{twoinvisible}
\end{figure}

By fixing $\epsilon=10^{-3}$ as an example, we plot the decay width as a function of the dark photon mass $m_U$ for both visible and invisible decay processes, as shown in Fig. \ref{twoinvisible}. 
We further select a set of typical dark photon masses $m_U=\{0.5, 1.0, 1.5, 2.0, 2.5\ \text{GeV}\}$, with the corresponding decay ratios $\Gamma/\Gamma_{J/\psi}$ listed in Table~\ref{two body invisible}.
It is evident from both the table and the curve that, for both visible and invisible decay channels, the decay widths the value declines sharply as $m_U$ approaches the $J/\psi$ threshold. The main reason for this curve trend is that $\Gamma$ contains denominator $1/[(m_U^2-M_{J/\psi}^2)+m_U^2\Gamma_U^2]$, and here the value of $\Gamma_U=\Gamma_{vis}+\Gamma_{inv}$ involving visible and invisible processes is relatively large, making $\Gamma$ directly proportional to $1/(m_U^2 \Gamma_U^2)$.
For all invisible decay channels combined, the expected event yield remains below single digits at BESIII, whereas it is about $0 \sim 47$ at STCF for masses below $ 0.8\ \text{GeV}$ with  $\epsilon=10^{-3}$. 
Strict management of statistical and systematic uncertainties is required as well.

\begin{figure}[!thbp]
    \centering   \includegraphics[width=0.6\textwidth]{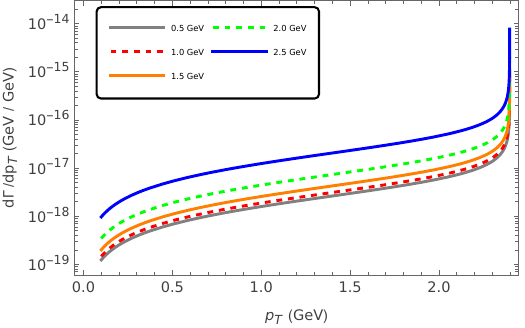}
    \caption{The differential cross-section distributions $d\sigma/dp_T$ for the process $J/\psi \to U \to e^+e^-$ with dark photon mass $m_U=\{0.5, \, 1.0, \, 1.5, \, 2.0, \, 2.5 \; {\rm GeV}\}$.} 
    \label{pt}
\end{figure}
\begin{table}[ht]
    \caption{Predictions for decay ratios $\Gamma/\Gamma_{J/\psi}$ of two body final state invisible decay process with $m_U=0.5, \, 1.0, \, 1.5, \, 2.0, \, 2.5, \, 3.0 \; {\rm GeV}$. The theoretical uncertainty entirely propagating from experimental error of $J/\psi$ mass. Here $\epsilon = 10^{-3}$ and $\alpha_\chi = 0.05$.}
    \begin{center}
    \renewcommand{\arraystretch}{1.4}
       \begin{tabular}{p{2.1cm}<{\centering}| p{3.3cm}<{\centering} |p{3.3cm}<{\centering}| p{3.3cm}<{\centering} | p{3.3cm}<{\centering} }
        \toprule
         \botrule
         $\mathcal{B} \, (\times 10^{-19})$ & $e^+e^-$  & $\mu^+\mu^-$  &  $\text{hadrons}$ & $\Gamma_{vis}/\Gamma_{J/\psi}$   \\
        \hline
$0.5 \; {\rm GeV}$ &  $1.0591 \pm 0.0191$ &  $1.0591 \pm 0.0191$ & $2.288 \pm 0.0412$ &  $4.406\pm0.0794$  \\
 \hline
$1.0 \; {\rm GeV}$ &  $0.0662 \pm 0.00119$ &  $0.0662 \pm 0.00119$ & $0.143 \pm 0.00258$ & $0.275\pm0.00496$   \\
 \hline
$1.5 \; {\rm GeV}$ &  $0.0131 \pm 0.000236$ &  $0.0131 \pm 0.000236$ & $0.0282 \pm 0.000509$ & $0.0544\pm0.000981$   \\
 \hline
$2.0 \; {\rm GeV}$ &  $0.00414 \pm 0.0000746$ &  $0.00414 \pm 0.0000746$ & $0.00894 \pm0.000161$ & $0.0172\pm0.000310$   \\    
 \hline
$2.5 \; {\rm GeV}$ &  $0.00169 \pm 0.0000305$ &  $0.00169 \pm 0.0000305$ & $0.00366 \pm0.0000660$ & $0.00705\pm0.000127$   \\
        \botrule
             $\mathcal{B} \, (\times 10^{-14})$& $\bar{\chi}_D\chi_D$  & $\bar{\chi}_M\chi_M$  &  $\varphi^\dagger \varphi$ & $\Gamma_{inv}/\Gamma_{J/\psi}$   \\
        \hline
$0.5 \; {\rm GeV}$ &  $72.580 \pm 1.308$ &  $71.326\pm 1.286$ & $35.663\pm0.643$ &  $179.569\pm3.237$  \\
 \hline
$1.0 \; {\rm GeV}$ &  $4.533 \pm 0.0817$ &  $4.225\pm 0.0762$ & $2.112\pm0.0381$ & $10.870\pm0.196$   \\
 \hline
$1.5 \; {\rm GeV}$ &  $0.892 \pm 0.0161$ &  $0.760\pm 0.0137$ & $0.380\pm0.00685$ & $2.0318\pm0.0366$   \\
 \hline
$2.0 \; {\rm GeV}$ &  $0.280 \pm 0.00504$ &  $0.208\pm 0.00376$ & $0.104\pm0.00188$ & $0.592\pm0.0107$   \\    
 \hline
$2.5 \; {\rm GeV}$ &  $0.112 \pm 0.00202$ &  $0.0695\pm 0.00125$ & $0.0348\pm0.000627$ & $0.216\pm0.00390$   \\
        \botrule
      \end{tabular}
    \end{center}
    \label{two body invisible}
\end{table}
%%%%%%%%%%%%%%%%%%%%%%%%%%%%%%%%%
\subsection{four-body final-state decay of $J/\psi$ mediated by a dark photon}
\subsubsection*{ $ {\rm (I)} \ m_U < 2m_\chi$}

\begin{figure}[!thbp]
    \centering   \includegraphics[width=0.75\textwidth]{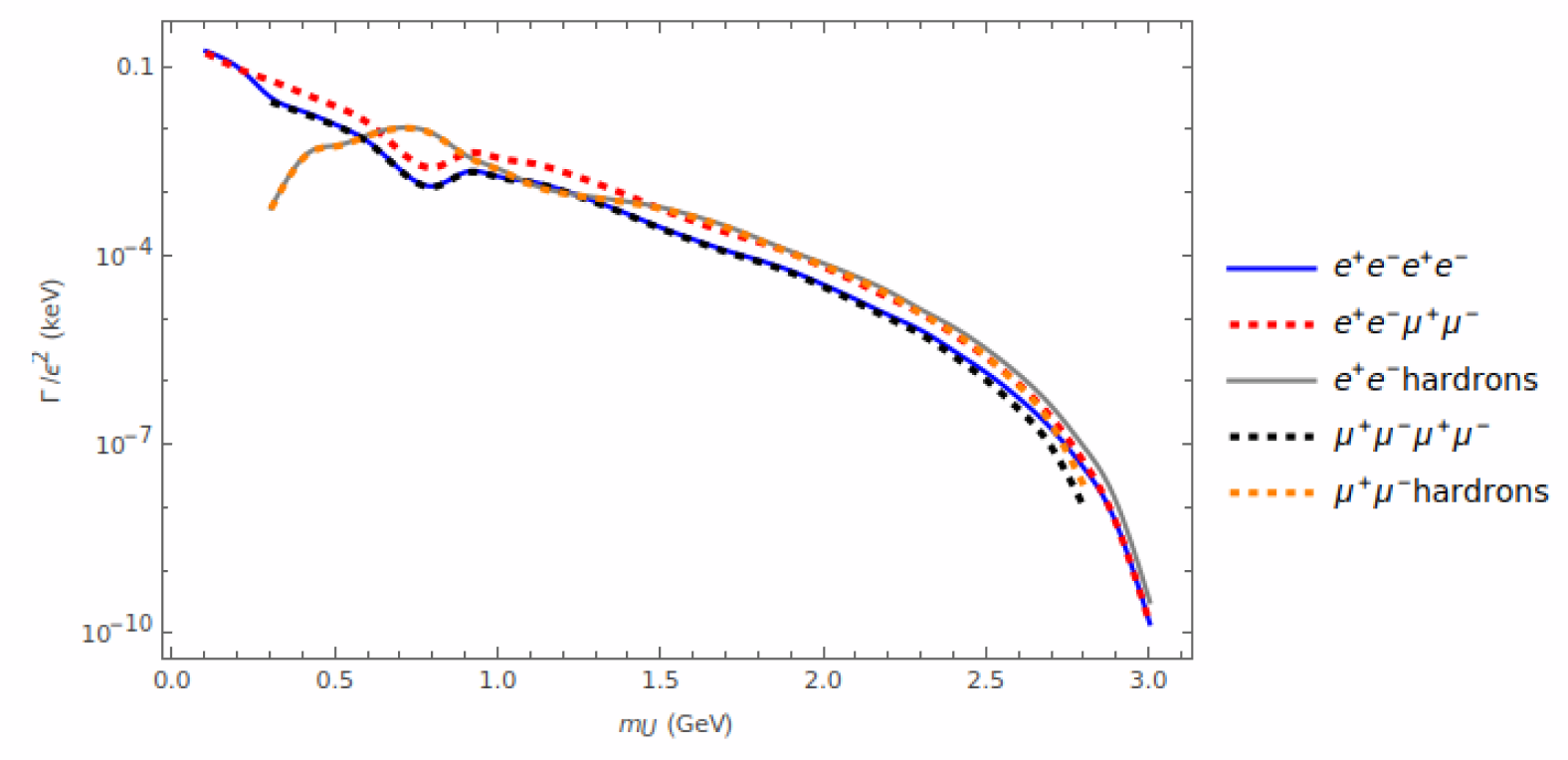}
    \caption{Decay width $\Gamma(J/\psi \to l^+l^-U \to l^+l^-X)$ divided by $\epsilon^2$ as a function of the dark photon mass $m_U$ for the four body final state decay processes $J/\psi \to l^+l^-U \to l^+l^-\mu^+\mu^-$, $J/\psi \to l^+l^-U \to l^+l^-e^+e^-$ and $J/\psi \to l^+l^-U \to l^+l^-{\rm hadrons}$.} 
    \label{fourbodyvisible}
\end{figure}

In Fig.~\ref{fourbodyvisible}, we show the decay width $\Gamma(J/\psi \to l^+l^-U \to l^+l^-X)$ divided by $\epsilon^2$ as a function of the dark photon mass $m_U$ for $l = e, \mu$ and $X = e^+e^-, \mu^+\mu^-, \text{hadrons}$. Here, $J/\psi \to e^+e^-\mu^+\mu^-$ includes $J/\psi \to e^+e^-U \to e^+e^-\mu^+\mu^-$ and $J/\psi \to \mu^+\mu^-U \to \mu^+\mu^-e^+e^-$. As shown in the figure, the magnitudes of all processes exhibit a roughly decreasing trend with $m_U$ up to $3.0\ \text{GeV}$, and the spectral shape of hadronic final-state channels is determined by $R(m_U)$. For masses lower than $2.2\ \text{GeV}$, signals become undetectable at BESIII when $\epsilon$ falls below $7.6\times10^{-5}$, and no detectable signals exist over the dark photon mass range $2.2\ \text{GeV} \leq m_U \leq 3.0\ \text{GeV}$. While STCF can push the corresponding $\epsilon$ limit down to $1.2\times10^{-5}$ within the full accessible mass range.

Then, we fix the coupling parameter $\epsilon = 10^{-4}$ as an example for the subsequent analysis. In Table~\ref{four body visible}, we present the decay ratio $\Gamma/\Gamma_{J/\psi}$ for typical masses of the dark photon $m_U=\{0.5, 1.0, 1.5, 2.0, 2.5\ \text{GeV}\}$. These results are obtained by applying the method outlined in Eq.~\eqref{B}. Based on the $J/\psi$ experimental data at STCF, the $e^+e^-e^+e^-$ and $e^+e^-\mu^+\mu^-$ final states can yield approximately $37\sim64$ and $35\sim62$ events for $m_U$ below $0.2\ \text{GeV}$, respectively, the corresponding the significance are $S/\sqrt{B}$ is $2.6\times10^{-3}\sim4.9\times10^{-3}$ and $3.1\times10^{-3}\sim5.9\times10^{-3}$, where the branching ratio of the background process are $\mathcal{B}(J/\psi \to e^+e^-e^+e^-)=5.5\pm 0.5 \times 10^{-5}$ and $\mathcal{B}(J/\psi \to e^+e^-\mu^+\mu^-)= 3.53\pm 0.26 \times 10^{-5}$ \cite{ParticleDataGroup:2024cfk}. For such a light dark photon, the only accessible decay channel is $e^+e^-$. In the mass range $0.2\ \text{GeV}$ to $3.0\ \text{GeV}$, the event yields from the final states with two lepton pairs become considerably smaller. 
For final state $l^+l^- + {\rm two \, jet}  $, the event yield is around 4 near $m_U=0.78\ \text{GeV}$, while being negligible at other masses. Nevertheless, all decay channels have event yields below single digits in BESIII.

\begin{table}[ht]
    \caption{Predictions for decay ratios $\Gamma/\Gamma_{J/\psi}$ of the four-body decay process $J/\psi \to l^+l^-U \to l^+l^-X$ , where $X$ denotes the visible decay products of $U$, with $m_U < 2m_\chi$ and $m_U=0.5, \, 1.0, \, 1.5, \, 2.0, \, 2.5 \; {\rm GeV}$. 
    The upper table shows the $J/\psi \to e^+e^-U$ intermediate state, and the lower table shows the $J/\psi \to \mu^+\mu^-U$ intermediate state. The theoretical uncertainty arises entirely from the $J/\psi$ mass experimental error. Here, $\epsilon = 10^{-4}$.}
    \begin{center}
    \renewcommand{\arraystretch}{1.4}
       \begin{tabular}{p{3.0cm}<{\centering}| p{2.3cm}<{\centering}| p{3.2cm}<{\centering} | p{3.2cm}<{\centering} | p{3.2cm}<{\centering} }
        \toprule
        \botrule
           & $\mathcal{B} \, (\times 10^{-14})$ & $e^+e^-e^+e^-$  & $e^+e^-\mu^+\mu^-$  &  $e^+e^-\text{hadrons}$    \\
        \hline
        \multirow{5}[0]{*}{$J/\psi \to e^+e^-X$} 
&$0.5 \; {\rm GeV}$ &  $132.990 \pm 2.398$ &  $131.271\pm 2.367$  &  $62.158\pm 1.121$  \\
 \cline{2-5}
&$1.0 \; {\rm GeV}$ &  $19.969 \pm 0.360$ &  $19.953\pm 0.360$  &  $26.384\pm 0.476$  \\
 \cline{2-5}
&$1.5 \; {\rm GeV}$ &  $3.0782 \pm 0.0555$ &  $3.0778\pm 0.0555$   &  $6.452\pm 0.116$  \\
 \cline{2-5}
&$2.0 \; {\rm GeV}$ &  $0.380 \pm 0.00685$ &  $0.380\pm 0.00685$  &  $0.828\pm 0.0149$   \\    
 \cline{2-5}
&$2.5 \; {\rm GeV}$ &  $0.0151 \pm 0.000272$ &  $0.0151\pm 0.000272$  &  $0.0360\pm 0.000649$  \\
\botrule
    &  &  $\mu^+\mu^-\mu^+\mu^-$ &  $\mu^+\mu^-e^+e^-$ &  $\mu^+\mu^-\text{hadrons}$  \\
\hline
\multirow{5}[0]{*}{$J/\psi \to \mu^+\mu^-X$} 
&$0.5 \; {\rm GeV}$ &  $127.531 \pm 2.299$ &  $129.201\pm 2.329$  &  $60.387\pm 1.0886$   \\
 \cline{2-5}
&$1.0 \; {\rm GeV}$ &  $19.390 \pm 0.350$ &  $19.405\pm 0.350$  &  $25.639\pm 0.462$  \\
\cline{2-5}
&$1.5 \; {\rm GeV}$ &  $2.953 \pm 0.0532$ &  $2.953\pm 0.0532$   &  $6.191\pm 0.112$  \\
\cline{2-5}
&$2.0 \; {\rm GeV}$ &  $0.350 \pm 0.00631$ &  $0.350\pm 0.00631$  &  $0.763\pm 0.0138$   \\    
 \cline{2-5}
&$2.5 \; {\rm GeV}$ &  $0.0114 \pm 0.000206$ &  $0.0114\pm 0.000206$  &  $0.0273\pm 0.000492$   \\
\botrule
      \end{tabular}
    \end{center}
    \label{four body visible}
\end{table}
%%%%%%%%%%%%%%%%%%%%%%%%%%%%%%%%%%%%%%%%%%%%%%%%%%%%%%%%%%%%%%%%%%%%%%%%%%%%%%%

\subsubsection*{ $ {\rm (II)} \ m_U \geq 2m_\chi$}

\begin{figure}[!thbp]
    \centering   \includegraphics[width=0.5\textwidth] {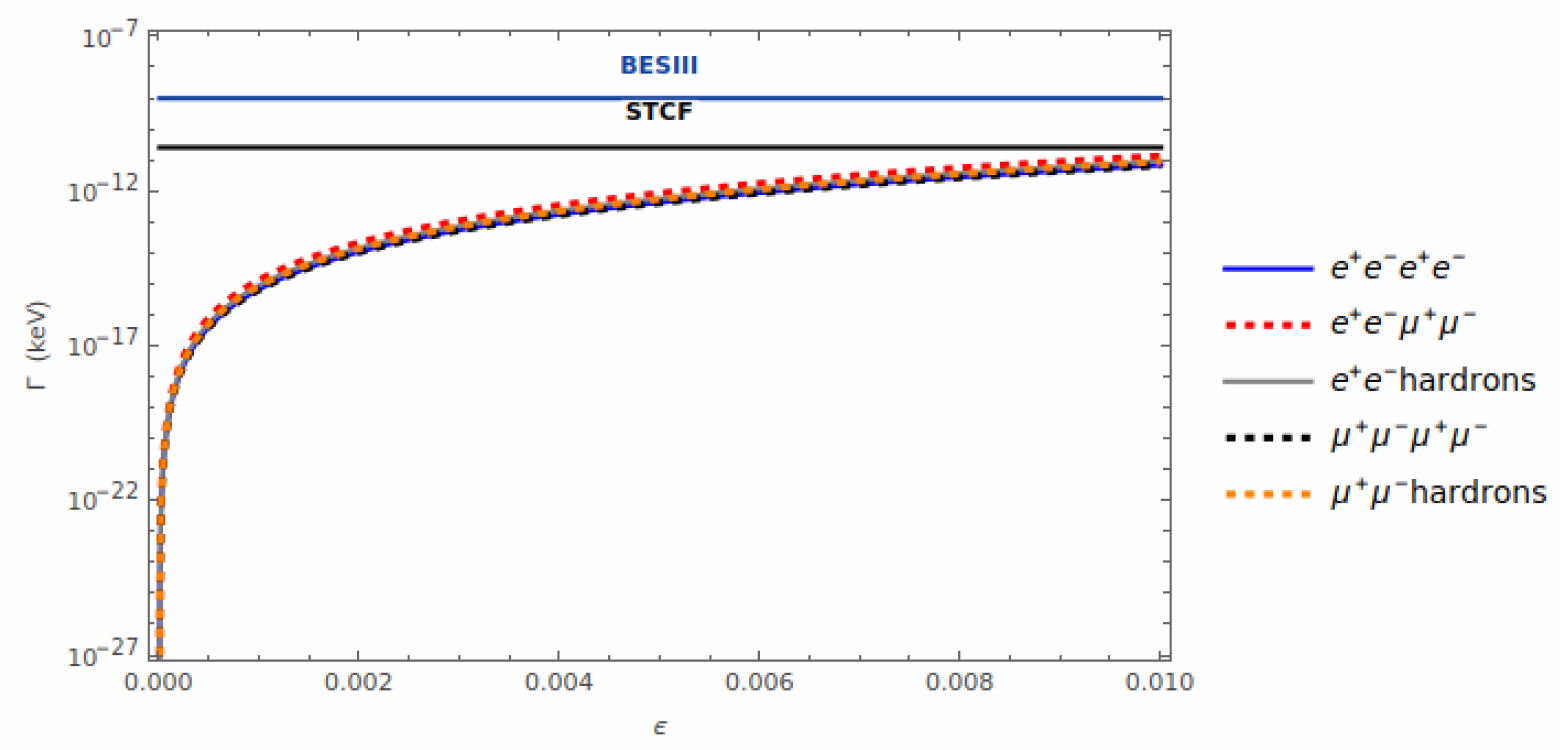}  \includegraphics[width=0.48\textwidth]{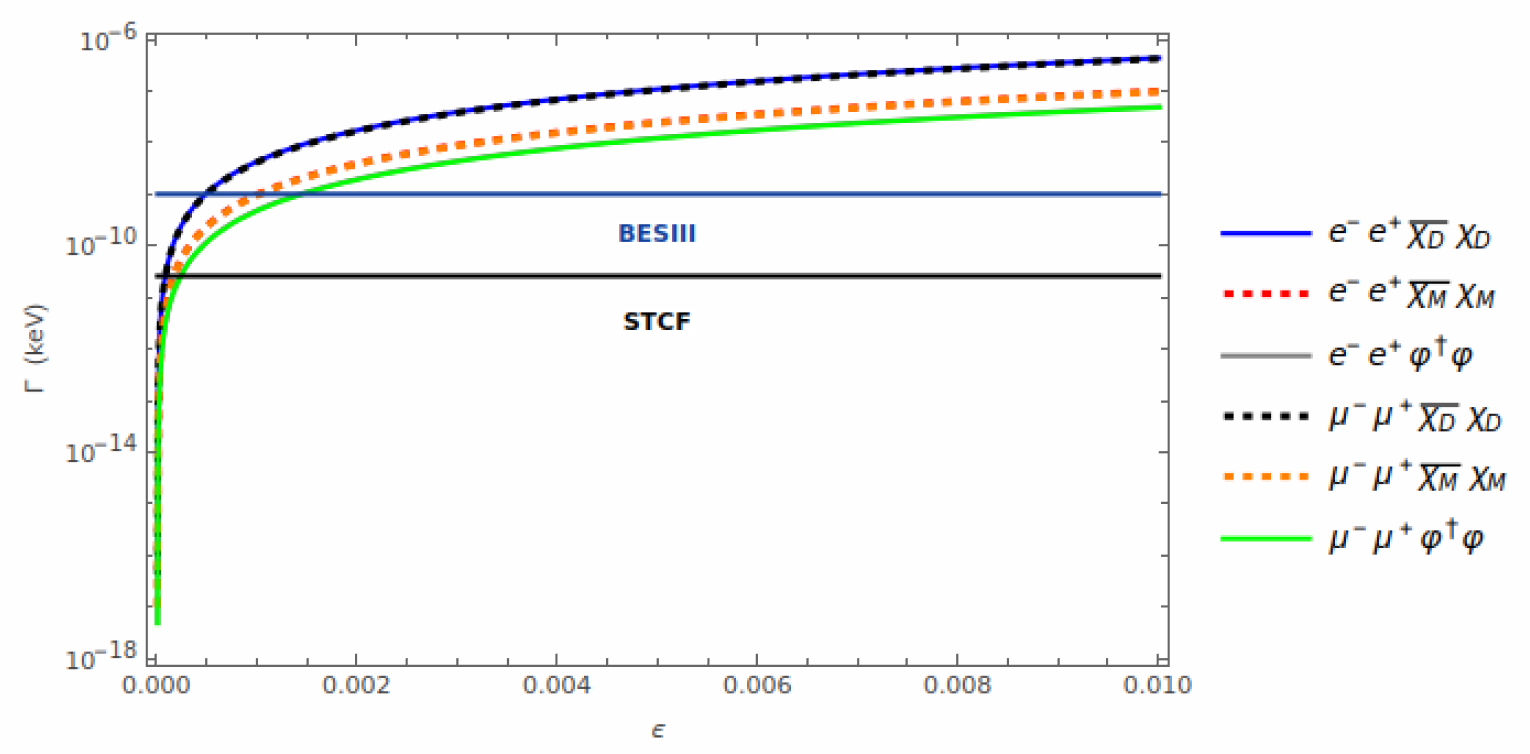}
    \caption{Left: Decay width as a function of the kinetic mixing parameter $\epsilon$ for the visible decay processes $J/\psi \to l^+l^-U \to l^+l^-\mu^+\mu^-$, $J/\psi \to l^+l^-U \to l^+l^-e^+e^-$ and $J/\psi \to l^+l^-U \to l^+l^-{\rm hadrons}$. The dark blue and black horizontal lines represent the experimental limits of the BESIII and the STCF, respectively. Right: Decay width as a function of the kinetic mixing parameter $\epsilon$ for the invisible decay processes $J/\psi \to l^+l^-U \to l^+l^-\bar{\chi}_D\chi_D$, $J/\psi \to l^+l^-U \to l^+l^-\bar{\chi}_M\chi_M$ and $J/\psi \to l^+l^-U \to l^+l^-\varphi^\dagger \varphi$, with $\alpha_\chi = 0.05$ and $m_U/m_\chi=3.0$. Both figures correspond to $m_U=1.0\; {\rm GeV}$.} 
    \label{epsilon3}
\end{figure}

In Fig.~\ref{epsilon3}, we display the decay width as a function of the kinetic mixing parameter $\epsilon$ for the four-body final state visible decay processes $J/\psi \to l^+l^-U \to l^+l^-\mu^+\mu^-$, $J/\psi \to l^+l^-U \to l^+l^-e^+e^-$ and $J/\psi \to l^+l^-U \to l^+l^-{\rm hadrons}$, as well as the invisible decay processes $J/\psi \to l^+l^-U \to l^+l^-\bar{\chi}_D\chi_D$, $J/\psi \to l^+l^-U \to l^+l^-\bar{\chi}_M\chi_M$ and $J/\psi \to l^+l^-U \to l^+l^-\varphi^\dagger \varphi$ by fixing the dark photon mass $m_U=1.0\; {\rm GeV}$ as an example. 
As the visible decay channels are strongly suppressed by invisible decay channels, all visible decay channels are almost excluded in both the BESIII and STCF experiments within the theoretical range of $\epsilon$.
For invisible decays, $\epsilon$ limit is $8.8\times10^{-5}$ for the dark photon mass below $2.4\ \text{GeV}$, but for masses range $2.4\ \text{GeV} \leq m_U \leq 3.0\ \text{GeV}$, no signals can be detected within the theoretical allowed range of $\epsilon$ in BESIII.
Meanwhile, the $\epsilon$ limit is $1.4\times10^{-5}$ for the dark photon mass range below $2.8\ \text{GeV}$, but for masses above $2.8\ \text{GeV}$, no signals can be detected in STCF.

\begin{figure}[!thbp]
    \centering   \includegraphics[width=0.5\textwidth] {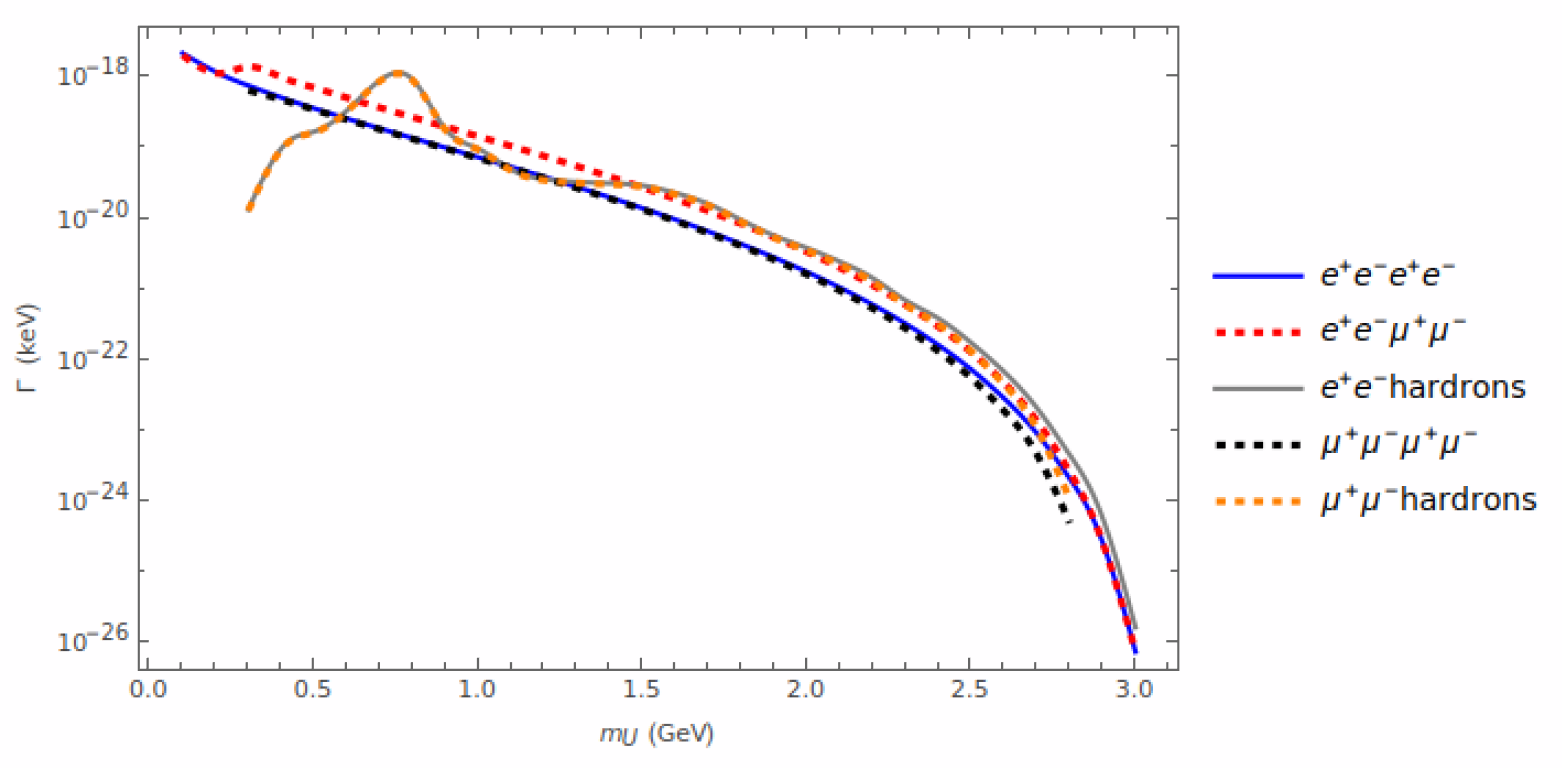}  \includegraphics[width=0.49\textwidth]{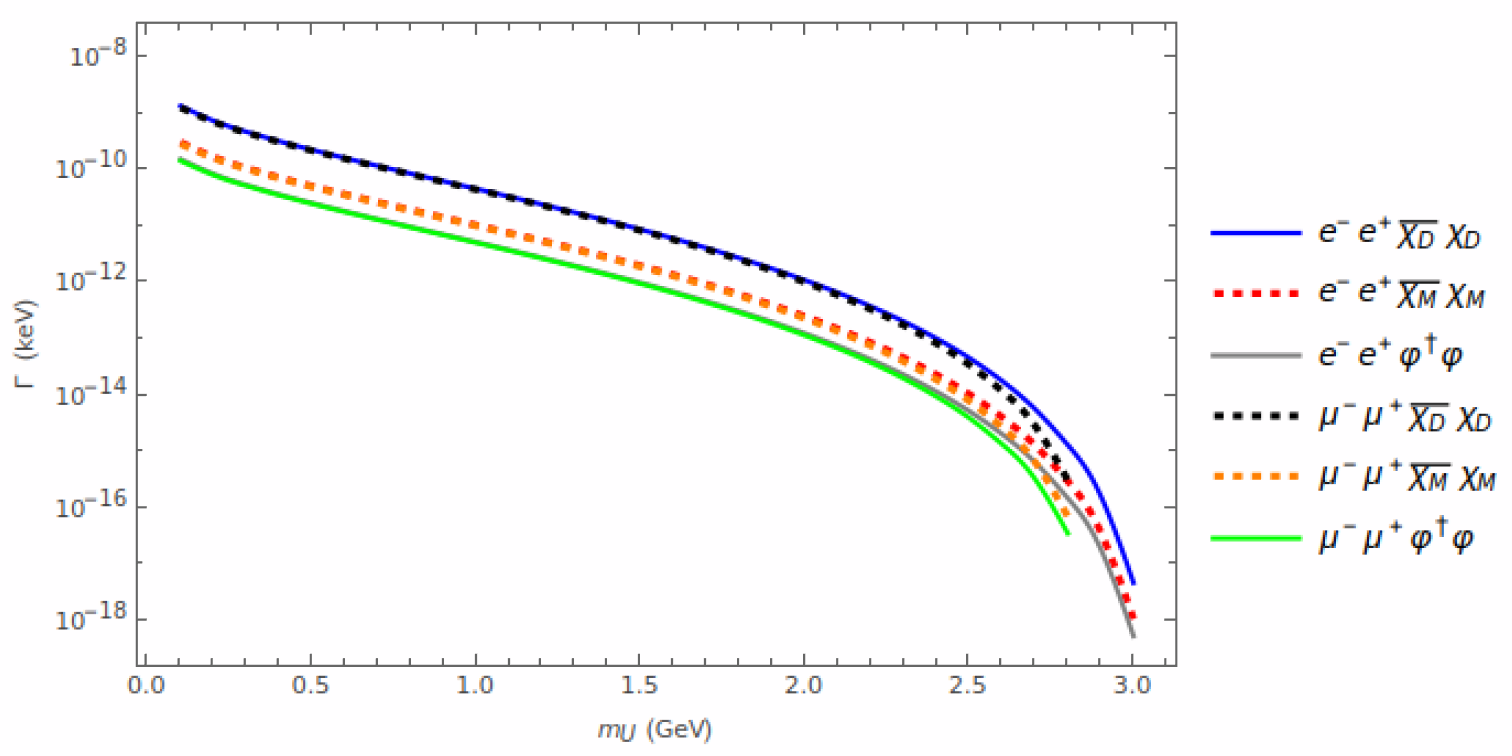}
    \caption{Left: Decay width as a function of the dark photon mass $m_U$ for the visible decay processes $J/\psi \to l^+l^-U \to l^+l^-\mu^+\mu^-$, $J/\psi \to l^+l^-U \to l^+l^-e^+e^-$ and $J/\psi \to l^+l^-U \to l^+l^-{\rm hadrons}$. Right: Decay width as a function of the dark photon mass $m_U$ for the invisible decay processes $J/\psi \to l^+l^-U \to l^+l^-\bar{\chi}_D\chi_D$, $J/\psi \to l^+l^-U \to l^+l^-\bar{\chi}_M\chi_M$ and $J/\psi \to l^+l^-U \to l^+l^-\varphi^\dagger \varphi$, with $\alpha_\chi = 0.05$ and $m_U/m_\chi=3.0$. Both figures correspond to $\epsilon=10^{-4}$.} 
    \label{fourbodyinvisible}
\end{figure}

By fixing the coupling parameter $\epsilon = 10^{-4}$ as an example, we evaluate the width of the four-body invisible decay process $J/\psi \to l^+l^-U \to l^+l^-X$ as a function of the dark photon mass $m_U$, where $l=e,\,\mu$ and $X=e^+e^-,\,\mu^+\mu^-,\,\text{hadrons},\,\bar{\chi}_D\chi_D,\,\bar{\chi}_M\chi_M,\,\varphi^\dagger\varphi$. Here we adopt $\alpha_\chi = 0.05$ and $m_U/m_\chi=3.0$, and present the results in Fig.~\ref{fourbodyinvisible}. As shown in the figure, both the visible decay width and the invisible decay width exhibit a roughly decreasing trend with increasing $m_U$. The variation trend of the curves for the four-body final state decay process is mainly dominated by process $J/\psi \to l^+l^-U$.
We also calculate the corresponding results for $m_U=\{0.5, 1.0, 1.5, 2.0, 2.5\ \text{GeV}\}$ and list them in Table~\ref{four body invisible} and Table~\ref{four body invisible2}.
The predicted event yields are below single digits at BESIII for the final state $l^+l^-+{\rm missing \, mass}$. At STCF, the corresponding yields are $0\sim 68$ and $0\sim 63$ for $l=e$ and $l=\mu$, respectively, when the dark photon mass is below $1.3\ \text{GeV}$. By contrast, even numbers for the SM particle final states are all negligible ($\ll 1$).
%%%

%%%%%
\begin{table}[ht]
    \caption{Predictions for decay ratios $\Gamma/\Gamma_{J/\psi}$ of the four-body decay process $J/\psi \to l^+l^-U \to l^+l^-X$ , where $X$ denotes the visible decay products of $U$, with $m_U \geq 2m_\chi$ and $m_U=0.5, \, 1.0, \, 1.5, \, 2.0, \, 2.5 \; {\rm GeV}$. The upper table shows the $J/\psi \to e^+e^-U$ intermediate state, and the lower table shows the $J/\psi \to \mu^+\mu^-U$ intermediate state. The theoretical uncertainty arises entirely from the $J/\psi$ mass experimental error. Here, $\epsilon = 10^{-4}$.}
    \begin{center}
    \renewcommand{\arraystretch}{1.4}
       \begin{tabular}{p{2.6cm}<{\centering}| p{2.3cm}<{\centering}| p{3.4cm}<{\centering} | p{3.4cm}<{\centering} | p{3.4cm}<{\centering} }
        \toprule
        \botrule
           & $\mathcal{B} \, (\times 10^{-21})$ & $e^+e^-e^+e^-$  & $e^+e^-\mu^+\mu^-$  &  $e^+e^-\text{hadrons}$  \\
        \hline
        \multirow{5}[0]{*}{$J/\psi \to e^+e^-X$} 
&$0.5 \; {\rm GeV}$ &  $3.899 \pm 0.0703$ &  $3.849\pm 0.0694$  &  $1.822\pm 0.0329$  \\
 \cline{2-5}
&$1.0 \; {\rm GeV}$ &  $0.792 \pm 0.0143$ &  $0.791\pm 0.0143$  &  $1.0465\pm 0.0189$ \\
 \cline{2-5}
&$1.5 \; {\rm GeV}$ &  $0.151 \pm 0.00272$ &  $0.151\pm 0.00271$   &  $0.316\pm 0.00569$ \\
 \cline{2-5}
&$2.0 \; {\rm GeV}$ &  $0.0190 \pm 0.000342$ &  $0.0190\pm 0.00342$  &  $0.0413\pm 0.000745$  \\    
 \cline{2-5}
&$2.5 \; {\rm GeV}$ &  $0.000790 \pm 0.0000142$ &  $0.000790\pm 0.0000142$  &  $0.00189\pm 0.0000340$ \\
\botrule
     & &  $\mu^+\mu^-\mu^+\mu^-$ &  $\mu^+\mu^-e^+e^-$ &  $\mu^+\mu^-\text{hadrons}$  \\
\hline
\multirow{5}[0]{*}{$J/\psi \to \mu^+\mu^-X$} 
&$0.5 \; {\rm GeV}$ &  $3.739 \pm 0.0674$ &  $3.788\pm 0.0683$  &  $1.770\pm 0.0319$  \\
 \cline{2-5}
&$1.0 \; {\rm GeV}$ &  $0.769 \pm 0.0139$ &  $0.770\pm 0.0139$  &  $1.0170\pm 0.0183$ \\
 \cline{2-5}
&$1.5 \; {\rm GeV}$ &  $0.144 \pm 0.00260$ &  $0.145\pm 0.00261$   &  $0.303\pm 0.00546$  \\
 \cline{2-5}
&$2.0 \; {\rm GeV}$ &  $0.0175 \pm 0.000315$ &  $0.0175\pm 0.000315$  &  $0.0381\pm 0.000687$   \\    
 \cline{2-5}
&$2.5 \; {\rm GeV}$ &  $0.000599 \pm 0.0000108$ &  $0.000599\pm 0.0000108$  &  $0.00143\pm 0.0000258$  \\
\botrule
      \end{tabular}
    \end{center}
    \label{four body invisible}
\end{table}

%%%%
\begin{table}[ht]
    \caption{Predictions for decay ratios $\Gamma/\Gamma_{J/\psi}$ of the four-body decay process $J/\psi \to l^+l^-U \to l^+l^-X$ , where $X$ denotes the invisible decay products of $U$, with $m_U \geq 2m_\chi$ and $m_U=0.5, \, 1.0, \, 1.5, \, 2.0, \, 2.5 \; {\rm GeV}$. The upper table shows the $J/\psi \to e^+e^-U$ intermediate state, and the lower table shows the $J/\psi \to \mu^+\mu^-U$ intermediate state. The theoretical uncertainty arises entirely from the $J/\psi$ mass experimental error. Here, $\epsilon = 10^{-4}$.}
    \begin{center}
    \renewcommand{\arraystretch}{1.4}
       \begin{tabular}{p{2.6cm}<{\centering}| p{2.3cm}<{\centering}| p{3.4cm}<{\centering} | p{3.4cm}<{\centering} | p{3.4cm}<{\centering} }
        \toprule
        \botrule
            &$\mathcal{B} \, (\times 10^{-14})$ & $e^+e^-\bar{\chi}_D\chi_D$  & $e^+e^-\bar{\chi}_M\chi_M$  &  $e^+e^-\varphi^\dagger \varphi$    \\
        \hline
        \multirow{5}[0]{*}{$J/\psi \to e^+e^-X$} 
&$0.5 \; {\rm GeV}$ &  $243.431 \pm 4.388$ &  $55.325\pm 0.997$  &  $27.663\pm 0.499$   \\
 \cline{2-5}
&$1.0 \; {\rm GeV}$ &  $49.449 \pm 0.891$ &  $11.238\pm 0.203$  &  $5.619\pm 0.101$ \\
 \cline{2-5}
&$1.5 \; {\rm GeV}$ &  $9.403 \pm 0.170$ &  $2.137\pm 0.0385$   &  $1.0685\pm 0.0193$  \\
 \cline{2-5}
&$2.0 \; {\rm GeV}$ &  $1.184 \pm 0.0213$ &  $0.269\pm 0.00485$  &  $0.135\pm 0.00243$  \\    
 \cline{2-5}
&$2.5 \; {\rm GeV}$ &  $0.0493 \pm 0.000889$ &  $0.0112\pm 0.000202$  &  $0.00561\pm 0.000101$ \\
\botrule
    &  &  $\mu^+\mu^-\bar{\chi}_D\chi_D$  & $\mu^+\mu^-\bar{\chi}_M\chi_M$  &  $\mu^+\mu^-\varphi^\dagger \varphi$  \\
\hline
\multirow{5}[0]{*}{$J/\psi \to \mu^+\mu^-X$} 
&$0.5 \; {\rm GeV}$ &  $236.495 \pm 4.263$ &  $53.749\pm 0.969$  &  $26.874\pm 0.484$   \\
 \cline{2-5}
&$1.0 \; {\rm GeV}$ &  $48.0532 \pm 0.866$ &  $10.921\pm 0.197$  &  $5.461\pm 0.0984$  \\
 \cline{2-5}
&$1.5 \; {\rm GeV}$ &  $9.0216 \pm 0.163$ &  $2.0504\pm 0.0370$   &  $1.0252\pm 0.0185$  \\
 \cline{2-5}
&$2.0 \; {\rm GeV}$ &  $1.0911 \pm 0.0197$ &  $0.248\pm 0.00447$  &  $0.124\pm 0.00224$   \\    
 \cline{2-5}
&$2.5 \; {\rm GeV}$ &  $0.0374 \pm 0.000674$ &  $0.00849\pm 0.000153$  &  $0.00425\pm 0.0000766$   \\
\botrule
      \end{tabular}
    \end{center}
    \label{four body invisible2}
\end{table}

%%%%%%%%%%%%%%%%%%%%%%%%%%%%%%%%%

%%%%%%%%%%%%%%%%%%%%%%%%%%%%%%%%%%%%%%%%%%
\clearpage
\section{SUMMARY}
We study $J/\psi$ decay processes mediated by the dark photon, which is introduced by an additional $U(1)_Y$ symmetry and kinetically mixes with the photon. The decays considered include two-body channels $J/\psi \to U \to X$ and four-body channels $J/\psi \to l^+l^-U \to l^+l^-X$. The final state $X$ contains only visible SM particles ($e^+e^-$, $\mu^+\mu^-$, or hadrons) when $m_U < 2m_\chi$. When $m_U \ge 2m_\chi$, $X$ includes both visible SM particles and invisible dark-sector particles (Dirac fermion pairs $\bar{\chi}_D\chi_D$, Majorana fermion pairs $\bar{\chi}_M\chi_M$, or complex scalar pairs $\varphi^\dagger\varphi$). Our study is conducted from two aspects concerning the influences of the dark photon mass $m_U$ and the kinetic mixing parameter $\epsilon$ on the results. Specifically, for the two-body final-state cases, we calculate the decay width versus $\epsilon$ and present the curves in Figs.~\ref{epsilon1} and Figs.~\ref{epsilon2} for $m_U= 1.0\ \text{GeV}$ as an example, the decay width versus $m_U$ and present the curves in Figs.~\ref{twovisiable} and Figs.~\ref{twoinvisible} with $\epsilon= 10^{-3}$ chosen as a benchmark, and the decay ratios $\Gamma/\Gamma_{J/\psi}$ in Tables~\ref{two body visible} and \ref{two body invisible} for typical masses $m_U=\{0.5, 1.0, 1.5, 2.0, 2.5\ \text{GeV}\}$. The differential transverse momentum distributions $d\sigma/dp_T$ for the $J/\psi \to U \to e^+e^-$ process are displayed in Fig.~\ref{pt}.
For the four-body final-state cases, we investigate the decay width $\Gamma(J/\psi \to l^+l^-U \to l^+l^-X)$ (see Eq.~\eqref{15}) divided by $\epsilon^2$ and present the curves in Figs.~\ref{fourbodyvisible} for $m_U < 2m_\chi$ condition. Meanwhile, for $m_U \geq 2m_\chi$ condition, we study the decay width versus $\epsilon$ and present the curves in Figs.~\ref{epsilon3} sitten $m_U= 1.0\ \text{GeV}$ for illustration, and the decay width as a function of $m_U$ with $\epsilon= 10^{-4}$ chosen as a benchmark, and present the results in Figs.~\ref{fourbodyinvisible}. The corresponding decay ratios $\Gamma/\Gamma_{J/\psi}$ are listed in Tables~\ref{four body visible} and \ref{four body invisible2}. All calculations are performed with $\alpha_\chi = 0.05$ and $m_U/m_\chi=3.0$. We investigate the sensitive parameter ranges of BESIII and STCF experiments with respect to parameters $\epsilon$ and $m_U$. The detailed analytical formulas are given in Appendix~\ref{Explicit Amplitude Form}.

Our results show that, for two-body final states with $m_U < 2m_\chi$, the expected upper limits on the kinetic mixing parameter $\epsilon$ at BESIII are approximately $9.3 \times 10^{-4}$ and $7.6 \times 10^{-4}$ for lepton-pair and hadron signals, respectively. The corresponding limits at STCF are $3.7 \times 10^{-4}$ and $3.1 \times 10^{-4}$, respectively.
When $m_U \geq 2m_\chi$, for invisible decays, the limit on $\epsilon$ is $1.4\times10^{-3}$ for $0.3\ \text{GeV} \leq m_U \leq 0.8\ \text{GeV}$ at BESIII, while the corresponding limit is $2.3\times10^{-4}$ over $0.3\ \text{GeV} \leq m_U \leq 1.9\ \text{GeV}$ at STCF. No signals can be detected for either experiment in other mass regions.  
For four-body final states decay channels, signals become undetectable for $m_U<2.2\ \text{GeV}$ when $\epsilon<7.6\times10^{-5}$, and no detectable signals exist in the range $2.2\ \text{GeV} \leq m_U \leq 3.0\ \text{GeV}$ at BESIII. For STCF, the corresponding $\epsilon$ limit can be pushed down to $1.2\times10^{-5}$ over the entire accessible mass range.
When $m_U \geq 2m_\chi$, visible decay channels nearly excluded at both BESIII and STCF within the theoretical $\epsilon$ range. For invisible decays, BESIII yields an $\epsilon$ limit of $8.8\times10^{-5}$ for $m_U<2.4\ \text{GeV}$, while no signals are detectable for $2.4\ \text{GeV} \leq m_U \leq 3.0\ \text{GeV}$. For STCF, the $\epsilon$ limit reaches $1.4\times10^{-5}$ for $m_U<2.8\ \text{GeV}$, and no signals can be detected at higher masses.
With the exception of parameter $9.3 \times 10^{-4}$, the above constraints on $\epsilon$ remain largely unexcluded by collider experiments. Compared with the limits from two-body final-state processes, the four-body decay channels yield constraints that are considerably lower than current experimental bounds. 

Nevertheless, for further phenomenological analysis and better comparison with experimental results, we fix the kinetic mixing parameter to $\epsilon=10^{-3}$ for two-body decay channels and $\epsilon=10^{-4}$ for two-body decay channels in our subsequent analysis. The value $\epsilon=10^{-3}$ has already been excluded by laboratory searches, while $\epsilon=10^{-4}$ is only ruled out at a few discrete mass points and remains allowed in our considered parameter space \cite{Caputo:2026pdw}. Therefore, our results can serve as an upper bound for two-body decay channels and provide supportive reference for constraining this parameter for BESIII and STCF via four-body decay channels.
\label{sec:summary}

\vspace{0.5cm}
\noindent\textbf{Acknowledgments}
We thank Jun Jiang and Wen-Yu Sun for helpful discussions on this work. This work is supported in part by the National Natural Science Foundation of China (12105162, 12305106, 12235008, 12321005), the Natural Science Foundation of Shandong Province (ZR2021QA058, ZR2021QA040) and the Youth Innovation Technology Project of Higher School in Shandong Province (2023KJ146).

%%%%%%%%%%%%%%%%%%%%%%%%%%%%%%%%%%%%%%%%%%
\clearpage
\appendix
\section{Explicit Form of expressions}

In this appendix, we list the explicit analytical expressions adopted throughout our calculations. Among them, $a$–$d$ correspond to the two-body decay widths of $J/\psi$; $e$ refers to the process $J/\psi \to l^+l^-U$; $f$–$i$ are for dark photon decays; and $j$ denotes the differential decay width of the process $J/\psi \to U \to l^+l^-$.

\label{Explicit Amplitude Form}

\paragraph{$\Gamma(J/\psi \to U \to l^+l^-)$}

\begin{equation}
    \begin{split} 
\frac{64 \, \pi \, \epsilon^4 \, \alpha^2 \, |\Psi_{J/\psi}(0)|^2 \, \sqrt{M_{J/\psi}^2-4 \, m_l^2}( M_{J/\psi}^2+ 2 \, m_l^2 ) }{9 \, M_{J/\psi} \, [(m_U^2 - M_{J/\psi}^2)^2+m_U^2\Gamma_{U}^2] }   \ .
    \end{split}
\end{equation}

\paragraph{$\Gamma(J/\psi \to U \to \bar{\chi}_D\chi_D)$}

\begin{equation}
    \begin{split} 
\frac{64 \, \pi  \, \alpha_\chi \, \epsilon^2 \,  \alpha \, |\Psi_{J/\psi}(0)|^2 \, \sqrt{M_{J/\psi}^2-4 \, m_\chi^2}( M_{J/\psi}^2+ 2 \, m_\chi^2 ) }{9 \, M_{J/\psi} \,  [(m_U^2 - M_{J/\psi}^2)^2+m_U^2\Gamma_{U}^2] }   \ .
    \end{split}
\end{equation}
\paragraph{$\Gamma(J/\psi \to U \to \bar{\chi}_M\chi_M)$}

\begin{equation}
    \begin{split} 
\frac{32 \,  \pi  \,   \alpha_\chi \, \epsilon^2 \,  \alpha \, |\Psi_{J/\psi}(0)|^2 \, (M_{J/\psi}^2-4 \, m_\chi^2)^{3/2}}{9 \, M_{J/\psi} \,  [(m_U^2 - M_{J/\psi}^2)^2+m_U^2\Gamma_{U}^2] }   \ .
    \end{split}
\end{equation}

\paragraph{$\Gamma(J/\psi \to U \to \varphi^\dagger \varphi)$}

\begin{equation}
    \begin{split} 
\frac{16 \, \pi  \, \alpha_\chi \, \epsilon^2 \,  \alpha \, |\Psi_{J/\psi}(0)|^2 \, (M_{J/\psi}^2-4 \, m_\chi^2)^{3/2}}{9 \, M_{J/\psi} \,  [(m_U^2 - M_{J/\psi}^2)^2+m_U^2\Gamma_{U}^2] }   \ .
    \end{split}
\end{equation}

\paragraph{$\Gamma(J/\psi \to l^+l^-U)$}

\begin{eqnarray}
\Gamma(J/\psi \to l^+l^-U) = \frac{1}{2 \, M_{J/\psi}} \int \frac{1}{3} \sum_{\rm spins}  \big|{\cal M}\big|^{2} d\Phi_3 \ ,
\label{sd-sigma}
\end{eqnarray}
where the squared amplitude $\big|{\cal M}\big|^{2}$ is 
\begin{equation}
    \begin{split} 
\big|{\cal M}\big|^{2} = &
  8192 \,\pi^3 \, \epsilon^2 \, \alpha^3 \,  |\Psi_{J/\psi}(0)|^2 
  \Bigl(
      -6 m_l^8
      + 2 m_U^4 (m_l^2 - s_2) (m_l^2 - s_3) 
      + s_2 s_3 \bigl(2 M_{J/\psi}^4 \\ & + s_2^2 +s_3^2 - 2 M_{J/\psi}^2 (s_2 + s_3)\bigr)
      + m_l^4 \bigl(2 M_{J/\psi}^4 + 3 s_2^2 + 14 s_2 s_3  + 3 s_3^2  - 2 M_{J/\psi}^2 \\ & (s_2 + s_3)\bigr)
      - m_l^2 \Bigl(-8 M_{J/\psi}^2 s_2 s_3 + 2 M_{J/\psi}^4 (s_2 + s_3) + (s_2 + s_3)(s_2^2 + 6 s_2 s_3 \\ & + s_3^2)\Bigr) 
      + m_U^2 \Bigl(
          2 m_l^4 (M_{J/\psi}^2 - s_2 - s_3)
          - 2 s_2 s_3 (s_2 + s_3)
          - M_{J/\psi}^2  (s_2^2 - 4 s_2 s_3  \\ & + s_3^2)
          - 2 m_l^2 \bigl(-4 s_2 s_3 + M_{J/\psi}^2  (s_2 + s_3)\bigr)
        \Bigr)
  \Bigr)
/
 9 \, M_{J/\psi}^3 (m_l^2 - s_2)^2 (m_l^2 - s_3)^2
   \ .
    \end{split}
\end{equation}

\paragraph{$\Gamma( U \to l^+l^-)$}

\begin{equation}
    \begin{split} 
\frac{ \epsilon^2 \,  \alpha \, \sqrt{m_U^2-4 \, m_l^2} \, (m_{U}^2+2 \, m_l^2)}{3 \, m_U^2 }   \ .
    \end{split}
\end{equation}

\paragraph{$\Gamma( U \to \bar{\chi}_D\chi_D)$}

\begin{equation}
    \begin{split} 
\frac{   \alpha_\chi \, \sqrt{m_U^2-4 \, m_\chi^2} \, (m_{U}^2+2 \, m_\chi^2)}{3 \, m_U^2 }   \ .
    \end{split}
\end{equation}

\paragraph{$\Gamma( U \to \bar{\chi}_M\chi_M)$}

\begin{equation}
    \begin{split} 
\frac{   \alpha_\chi \, (m_U^2-4 \, m_\chi^2)^{3/2}}{6 \, m_U^2 }   \ .
    \end{split}
\end{equation}

\paragraph{$\Gamma( U \to \varphi^\dagger \varphi)$}

\begin{equation}
    \begin{split} 
\frac{   \alpha_\chi \, (m_U^2-4 \, m_\chi^2)^{3/2}}{12 \, m_U^2 }   \ .
    \end{split}
\end{equation}

\paragraph{$d\Gamma( J/\psi \to U \to l^+l^-)/dp_T$}

\begin{equation}
    \begin{split} 
\frac{128 \, \pi \, \epsilon^4 \, \alpha^2 \, |\Psi_{J/\psi}(0)|^2 \, ( M_{J/\psi}^2+ 2 \, m_l^2 ) \, p_T }{9 \, M_{J/\psi} \, [(m_U^2 - M_{J/\psi}^2)^2+m_U^2\Gamma_{U}^2] \, \sqrt{M_{J/\psi}^2-4 \, m_l^2-4 \, p_T} }   \ .
    \end{split}
\end{equation}

%%%%%%%%%%%%%%%%%%%%%%%%%%%%%%%%%%%%%%%%%%%%%%%%%%%%%%%%%%%%%%%%%%%%%
%%%%%%%%%%%%%%%%%%%%%%%%%%%%%%%%%%%%%%%%%%%%%%%%%%%%%%%%%%%%%%

%%%%%%%%%%%%%%%%%%%%%%%%%%%%%%%%%%%%%%%%%%%%%%%%%%%%%%%%%%%%%%%%%%%%%%
%\include{ref.tex}

\end{document}